\documentclass[aps,prd,reprint,superscriptaddress,showpacs,floatfix,nofootinbib]{revtex4-2}  
\usepackage[hmargin=1.5cm,top=1.5cm,bottom=2cm]{geometry}
\usepackage{graphicx}
\usepackage{amsmath,amssymb,graphics,setspace}
\usepackage{color}
\usepackage[english]{babel}
\usepackage{url,bm}
\usepackage{amsfonts}
\usepackage[separate-uncertainty = true]{siunitx}
\usepackage{dcolumn}
\usepackage{appendix}
\usepackage{lmodern}
\usepackage{microtype}
\usepackage{esdiff}
\usepackage[table,dvipsnames]{xcolor}
\usepackage[babel]{csquotes}
\usepackage{textcomp}
\usepackage{listings}
\usepackage{enumitem}
\usepackage{lipsum}
\usepackage{tabularx}
\usepackage{makecell}
\usepackage{soul}

\newcommand{\ORCIDiD}{\includegraphics[width=2ex]{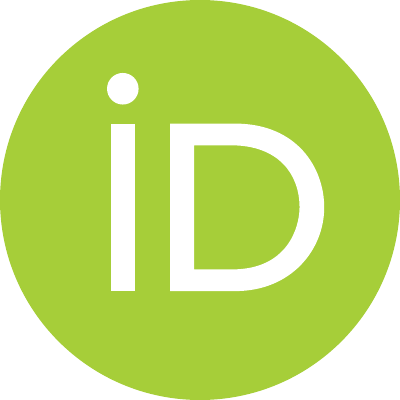}}
\usepackage{tikz}
\usetikzlibrary{tikzmark}

\usepackage[hidelinks]{hyperref}
\usepackage{natbib}
\usepackage{wasysym}

\newcommand{\mathsym}[1]{{}}
\newcommand{\unicode}[1]{{}}
\newcommand{\rtHz}{\ensuremath{\sqrt{\text{Hz}}}}

\newcommand{\OMS}{\text{OMS}}
\newcommand{\GRS}{\text{GRS}}

\newcommand{\dgmax}{\Delta g_\text{max}}
\DeclareSIUnit\year{yr}

\definecolor{HotPink}{RGB}{255 ,0,128}
\definecolor{cardinal}{rgb}{0.77, 0.12, 0.23}
\definecolor{Gorange}{RGB}{255,131,36}
\definecolor{lightgray}{gray}{0.85}
{}

\setstcolor{SeaGreen}

\newcommand{\numonesided}{432 } 
\newcommand{\numonesidedcontinuity}{384 }
\newcommand{\numonesidedtwotau}{48 }
\newcommand{\numtwosidedCR}{152 } 
\newcommand{\numtwosidedOR}{4 } 
\makeatletter
\renewcommand*{\@fnsymbol}[1]{\ensuremath{\ifcase#1\or \dagger\or a \or b \or c \or d \or e \or ** \else\@ctrerr\fi}}
\makeatother

\begin{document}
\title{Transient acceleration events in LISA Pathfinder data:\texorpdfstring{\\}{}
Properties and possible physical origin}
\date{\today}

\author{M~Armano}\affiliation{\addressa}
\author{H~Audley}\affiliation{\addressb}
\author{J~Baird}\affiliation{\addressca}
\author{P~Binetruy}\thanks{Deceased.}\affiliation{\addressc} 
\author{M~Born}\affiliation{\addressb}
\author{D~Bortoluzzi}\affiliation{\addressf}
\author{E~Castelli~\href{https://orcid.org/0000-0002-4429-0682}{\ORCIDiD}}\altaffiliation[Current address: ]{\addressu}
\affiliation{\addressi}
\author{A~Cavalleri}\affiliation{\addressk}
\author{A~Cesarini}\affiliation{\addresso}
\author{V\,Chiavegato}\affiliation{\addressi}
\author{A\,M~Cruise}\affiliation{\addressj}
\author{D\,Dal Bosco}\affiliation{\addressi}
\author{K~Danzmann}\affiliation{\addressb}
\author{M~De Deus Silva}\affiliation{\addressa}
\author{I~Diepholz}\affiliation{\addressb}
\author{G~Dixon}\affiliation{\addressj}
\author{R~Dolesi}\affiliation{\addressi}
\author{L~Ferraioli}\affiliation{\addressl}
\author{V~Ferroni}\affiliation{\addressi}
\author{E\,D~Fitzsimons}\affiliation{\addressm}
\author{M~Freschi}\affiliation{\addressa}
\author{L~Gesa}\thanks{Deceased.}\affiliation{\addressn} 
\author{D~Giardini}\affiliation{\addressl}
\author{F~Gibert}\altaffiliation[Current address: ]{\addressee}\affiliation{\addressi}
\author{R~Giusteri}\affiliation{\addressb}
\author{C~Grimani}\affiliation{\addresso}
\author{J~Grzymisch}\affiliation{\addressh}
\author{I~Harrison}\affiliation{\addressp}
\author{M\,S~Hartig}\affiliation{\addressb}
\author{G~Heinzel}\affiliation{\addressb}
\author{M~Hewitson}\affiliation{\addressb}
\author{D~Hollington}\affiliation{\addressd}
\author{D~Hoyland}\affiliation{\addressj}
\author{M~Hueller}\affiliation{\addressi}
\author{H~Inchausp\'e}\altaffiliation[Current address: ]{\addresscb}\affiliation{\addressca}
\author{O~Jennrich}\affiliation{\addressh}
\author{P~Jetzer}\affiliation{\addressq}
\author{B~Johlander}\affiliation{\addressa}
\author{N~Karnesis}\affiliation{\addressbb}
\author{B~Kaune}\affiliation{\addressb}
\author{N~Korsakova}\affiliation{\addressca}
\author{C\,J~Killow}\affiliation{\addressr}
\author{J\,A~Lobo}\thanks{Deceased.}\affiliation{\addressn} 
\author{J\,P~L\'opez-Zaragoza}\affiliation{\addressn}
\author{R~Maarschalkerweerd}\affiliation{\addressp}
\author{D~Mance}\affiliation{\addressl}
\author{V~Mart\'{i}n}\affiliation{\addressn}
\author{L~Martin-Polo}\affiliation{\addressa}
\author{F~Martin-Porqueras}\affiliation{\addressa}
\author{J~Martino}\affiliation{\addressca}
\author{P\,W~McNamara}\affiliation{\addressh}
\author{J~Mendes}\affiliation{\addressp}
\author{L~Mendes}\affiliation{\addressa}
\author{N~Meshksar}\affiliation{\addressl}
\author{M~Nofrarias}\affiliation{\addressn}
\author{S~Paczkowski}\affiliation{\addressb}
\author{M~Perreur-Lloyd}\affiliation{\addressr}
\author{A~Petiteau}\affiliation{\addressc}\affiliation{\addressca}
\author{E~Plagnol}\affiliation{\addressca}
\author{J~Ramos-Castro}\affiliation{\addresss}
\author{J~Reiche}\affiliation{\addressb}
\author{F~Rivas}\altaffiliation[Current address: ]{\addresscc}\affiliation{\addressi}
\author{D\,I~Robertson}\affiliation{\addressr}
\author{G~Russano}\altaffiliation[Current address: ]{\addressx}\affiliation{\addressi}
\author{L~Sala~\href{https://orcid.org/0000-0002-2682-8274}{\ORCIDiD}}\affiliation{\addressi}
\author{P~Sarra}\affiliation{\addressaa}
\author{J~Slutsky}\affiliation{\addressu}
\author{C\,F~Sopuerta}\affiliation{\addressn}
\author{T~Sumner}\affiliation{\addressd}
\author{D~Texier}\affiliation{\addressa}
\author{J\,I~Thorpe}\affiliation{\addressu}
\author{D~Vetrugno}\affiliation{\addressi}
\author{S~Vitale~\href{https://orcid.org/0000-0002-2427-8918}{\ORCIDiD}}\affiliation{\addressi}
\author{G~Wanner}\affiliation{\addressb}
\author{H~Ward}\affiliation{\addressr}
\author{P~Wass}\affiliation{\addressd}\affiliation{\addressdd}
\author{W\,J~Weber}\affiliation{\addressi}
\author{L~Wissel}\affiliation{\addressb}
\author{A~Wittchen}\affiliation{\addressb}
\author{C~Zanoni}\affiliation{\addressi}
\author{P~Zweifel}\affiliation{\addressl}

\collaboration{LISA Pathfinder Collaboration}\email[Corresponding authors:\\  ]{eleonora.castelli@nasa.gov\\lorenzo.sala@unitn.it\\stefano.vitale@unitn.it}

\def\addressa{European Space Astronomy Centre, European Space Agency, Villanueva de la
Ca\~{n}ada, 28692 Madrid, Spain}
\def\addressb{Albert-Einstein-Institut, Max-Planck-Institut f\"ur Gravitationsphysik und Leibniz Universit\"at Hannover,
Callinstra{\ss}e 38, 30167 Hannover, Germany}
\def\addressc{IRFU, CEA, Universit\'e Paris-Saclay, F-91191 Gif-sur-Yvette, France}
\def\addressca{Universit\'e Paris Cit\'e, CNRS, Astroparticule et Cosmologie, F-75013 Paris, France}
\def\addresscb{Institut f\"ur Theoretische Physik, Universit\"at Heidelberg, Philosophenweg 16, 69120 Heidelberg, Germany}
\def\addressd{Physics Department, Blackett Laboratory, High Energy Physics Group, Imperial College London, Prince Consort Road, London SW7 2BW, United Kingdom}
\def\addresse{Dipartimento di Fisica, Universit\`a di Roma ``Tor Vergata'',  and INFN, sezione Roma Tor Vergata, I-00133 Roma, Italy}
\def\addressf{Department of Industrial Engineering, Universit\`a di Trento and Trento Institute for 
Fundamental Physics and Application / INFN, 38123 Povo, Trento, Italy}
\def\addressg{Airbus Defence and Space, Claude-Dornier-Strasse, 88090 Immenstaad, Germany}
\def\addressh{European Space Technology Centre, European Space Agency, Keplerlaan 1, 2200 AG Noordwijk, Netherlands}
\def\addressi{Dipartimento di Fisica, Universit\`a di Trento and Trento Institute for 
Fundamental Physics and Application / INFN, 38123 Povo, Trento, Italy}
\def\addressj{The School of Physics and Astronomy, University of
Birmingham, B15 2TT Birmingham, United Kingdom}
\def\addressk{Istituto di Fotonica e Nanotecnologie, CNR-Fondazione Bruno Kessler, 
    I-38123 Povo, Trento, Italy}
\def\addressl{Institut f\"ur Geophysik, ETH Z\"urich, Sonneggstrasse 5, CH-8092 Z\"urich, Switzerland}
\def\addressm{The UK Astronomy Technology Centre, Royal Observatory, Edinburgh, Blackford Hill, Edinburgh EH9 3HJ, United Kingdom}
\def\addressn{Institut de Ci\`encies de l'Espai (CSIC-IEEC), Campus UAB, Carrer de Can Magrans s/n, 08193 Cerdanyola del Vall\`es, Spain}
\def\addresso{DISPEA, Universit\`a di Urbino Carlo Bo, Via S. Chiara, 27 61029 Urbino/INFN, Italy}
\def\addressp{European Space Operations Centre, European Space Agency, 64293 Darmstadt, Germany }
\def\addressq{Physik Institut, 
Universit\"at Z\"urich, Winterthurerstrasse 190, CH-8057 Z\"urich, Switzerland}
\def\addressr{SUPA, Institute for Gravitational Research, School of Physics and Astronomy, University of Glasgow, Glasgow G12 8QQ, United Kingdom}
\def\addresss{Department d'Enginyeria Electr\`onica, Universitat Polit\`ecnica de Catalunya,  08034 Barcelona, Spain}
\def\addressu{Gravitational Astrophysics Lab, NASA Goddard Space Flight Center, 8800 Greenbelt Road, Greenbelt, Maryland 20771 USA}
\def\addressx{INAF Osservatorio Astronomico di Capodimonte, I-80131 Napoli, Italy}
\def\addressy{INFN - Sezione di Napoli, I-80126, Napoli, Italy}
\def\addressz{Dipartimento di Fisica ed Astronomia, Universit\`a degli Studi di Firenze and INFN - Sezione di Firenze, I-50019 Firenze, Italy}
\def\addressaa{OHB Italia S.p.A, Via Gallarate, 150 - 20151 Milano, Italy}
\def\addressbb{Department of Physics, Aristotle University of Thessaloniki, Thessaloniki 54124, Greece}
\def\addresscc{Universidad Loyola Andaluc\'ia, Department of Quantitative Methods, Avenida de las Universidades s/n, 41704, Dos Hermanas, Sevilla, Spain}
\def\addressdd{Department of Mechanical and Aerospace Engineering, MAE-A, P.O. Box 116250, University of Florida, Gainesville, Florida 32611, USA}
\def\addressee{isardSAT SL, Marie Curie 8-14, 08042 Barcelona, Catalonia, Spain}

\begin{abstract}
We present an in depth analysis of the transient events, or glitches, detected at a rate of about one per day in the differential acceleration data of LISA Pathfinder. We show that these glitches fall in two rather distinct categories: fast  transients in the interferometric motion readout on one side, and true force transient events on the other. The former are fast and  rare in ordinary conditions. The second may last from seconds to hours and constitute the majority of the glitches.
We present an analysis of the physical and statistical properties of both categories, including a cross-analysis with  other time series like magnetic fields, temperature, and other dynamical variables.
Based on these analyses we discuss  the possible sources of the force glitches and identify the most likely, among which  the outgassing environment surrounding the test-masses stands out. We discuss the impact of these findings on the LISA design and operation, and  some risk mitigation measures, including experimental studies that may be conducted on the ground, aimed at clarifying some of the questions left open by our analysis.
\end{abstract}

\renewcommand{\figurename}{FIG.} 
\renewcommand{\tablename}{TABLE} 

\interfootnotelinepenalty=10000
\maketitle
\section{Introduction} 
\label{sec:introduction}
The European Space Agency (ESA) launched and operated the LISA Pathfinder (LPF) mission \cite{PhysRevLett.120.061101,armano:subfemtog} between December 2015 and July 2017. The scientific goal of LPF was to demonstrate that parasitic forces on a test mass, to be used as a geodesic reference in the future LISA gravitational wave observatory \cite{Amaro-Seoane2017}, may be suppressed below the required level. 

To that aim, the mission carried a miniature version of one of the LISA interferometric arms, that is, two kilogram-size free-orbiting test masses, separated by a few tens of centimeters, and an interferometric readout measuring their relative acceleration along the line joining their respective centers of mass.

The mission surpassed its goals and found that the acceleration due to parasitic forces had a power spectral density (PSD) better than LISA requirements across the entire observational frequency band $[\SI{20}{\micro\hertz},\SI{1}{\hertz}]$  \cite{PhysRevLett.120.061101}. 

Acceleration noise was found to be stationary over the week-long measurement runs, allowing a consistent PSD estimation. In addition, the measured PSD was found to be rather stable over the more than one year  duration of the mission science operations, except for the decrease of the Brownian noise following the corresponding decrease of the pressure surrounding the test masses \cite{PhysRevLett.120.061101}.

Besides this quasistationary noise, the acceleration data series also contained isolated events with different shapes and amplitudes which we call \emph{glitches}. These events are \emph{de facto} signals and can be modeled and subtracted from the data, a procedure which is essential, at least for the most energetic ones, to get a consistent estimate of the PSD of the underlying quasistationary noise over the entire data series.  

The purpose of this paper is to present a comprehensive description of glitches, an analysis of their physical properties, and finally a discussion of the possible physical sources. The paper is organized as follows. In Sec.~\ref{sec:exp}, we show the experimental layout and the measured data series. In Sec.~\ref{sec:gltchdetection}, we provide a description of the glitch detection and parameter estimation techniques. In  Sec.~\ref{sec:glitchparamstats}, we give the main features of glitch parameter  statistics. In Sec.~\ref{sec:jointanalysis}, we present the results of a coincidence analysis with other auxiliary data series. Finally, in Sec.~\ref{sec:disc}, we describe possible mechanisms for glitch generation and the likelihood of each. The mechanisms include (a) platform accelerations, (b) thermal effects, (c) gravitational signals, (d) magnetic forces, (e) electrostatic forces, and (f) outgassing. We discuss each possible source, and, while there is no conclusive evidence, find that outgassing events could lead to signals with similar properties to the measured ones. In the conclusions, Sec.~\ref{sec:conclusion}, we discuss the implications of these findings for the LISA mission.

\section{\label{sec:exp}Summary description of the experiment} 
\subsection{The LISA Technology Package}
The instrument flown on LPF, the LISA Technology Package (LTP), has been described in detail in \cite{LPFPKS}. Here we summarize its essential features.

\begin{figure}[htbp]
  \centering
  \includegraphics[width=\columnwidth]{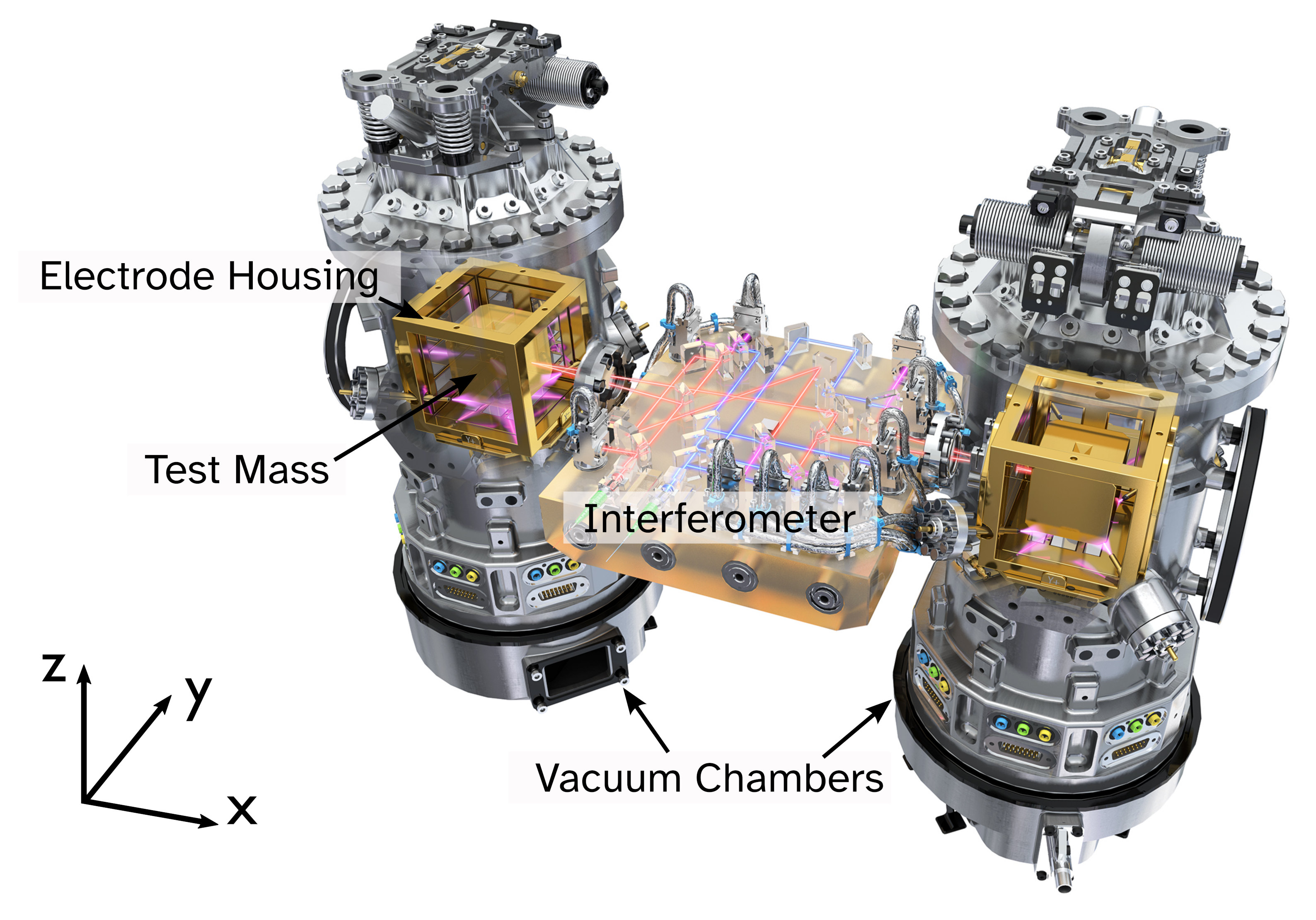}
  \caption{Rendering of the  LISA Technology Package. The rendering shows the two test-masses hosted inside their respective electrode-housings (some of the electrodes are not represented), and the vacuum chambers enclosing both test-masses and electrode housings. The picture also shows the high stability optical bench hosting all interferometric readouts, and many other features of the instrument, launch lock, UV-light based test-mass neutralizer, etc. that are not relevant here. [Copyright: ESA/ATG medialab]}
  \label{fig:LPFfigure}
\end{figure}

The LTP, depicted in Fig.~\ref{fig:LPFfigure}, carried two cubic Au-Pt test masses each with a mass of \SI{1.928}{kg} and size \SI{46}{mm}. During operation these test masses had no mechanical contact with their surroundings, and were \emph{free falling}, each one at the center of a  housing leaving 3 to \SI{4}{mm} clearance gaps to the faces of the test mass. Each of these \emph{electrode housings} carried a series of electrodes facing all faces of the respective test mass. These electrodes were used for two purposes. First they were part of a capacitive sensor of the motion of the test mass relative to its housing, for all  degrees of freedom. Second, they were used to apply feedback electrostatic forces and torques to the test mass, whenever needed.

The main sensor for the relative motion of the test masses was a heterodyne laser interferometric system, called the optical metrology system (OMS)  \cite{PhysRevLett.126.131103}. For the purpose of this paper it is important to recall that the interferometer measured six different degrees of freedom: the relative displacement $\Delta x(t)$ between the test masses along the sensitive $x$-axis, joining their respective centers of mass; the relative displacement $x_1(t)$ along the $x$-axis of one of the test mass, called TM1, relative to the interferometer optical bench and, as a consequence, also relative to the spacecraft; the two angles of rotation $\eta (t)$ and  $\phi(t)$ for both test masses, around the $y$-axis and $z$-axis respectively. These six test mass degrees of freedom, and also the remaining six, were also measured at all times by the capacitive sensors \cite{PRD_96_062004_capacitive}. However the interferometric readout was approximately three orders of magnitude more sensitive than the capacitive one for all degrees of freedom for which they were both available. All the measurements above have a sampling rate of \SI{10}{\hertz}.

Each electrode housing, with its respective test mass, was hosted inside a vacuum chamber that was vented to space via a dedicated valve. The chamber was needed to handle the vacuum on ground, and  because the outgassing within the spacecraft once in orbit was too large to achieve the desired vacuum level around the test masses. Thus the chamber was evacuated and sealed on ground and then, once on orbit, vented to space via a dedicated duct. In what follows we call the gravitational reference sensor (GRS) the system of the test mass, its electrode housing and vacuum chamber, and all related accessories.

Besides the measurement of the test mass motion, other physical quantities have been measured throughout the mission. In particular  we measured: the magnetic field vector at various locations, via a dedicated set of magnetometers \cite{magnetic-mnras}; the temperature at various critical locations,  via a dedicated set of thermistors  \cite{10.1093/mnras/stz1017}; the cosmic ray flux with a radiation monitor \cite{ARMANO201828,armano_characteristics_2018}; finally, two additional interferometric readouts, one monitoring frequency fluctuations and the other one monitoring common mode noise sources as a reference \cite{PhysRevLett.126.131103}. The reference interferometer time series has therefore been subtracted from the $\Delta x$, $x_1$ and frequency interferometer time series.

\subsection{\label{sec:dyn} Dynamical controls and data series formation}
LPF was a controlled dynamical system consisting of the spacecraft and the two test masses. More specifically, the spacecraft was forced to follow one of the test masses (TM1) along $x$ via an active control loop, using the spacecraft cold gas microthrusters as actuators \cite{PhysRevD.99.082001}, known as drag-free control. 

The test mass rotation along $\phi$ and $\eta$ was kept fixed relative to the spacecraft by an active loop using electrostatic torques. These torques were applied via the above mentioned electrodes.

No force was applied along $x$ on TM1, while a control loop (electrostatic suspension) kept the distance between the two test masses nominally fixed, by applying a suitable electrostatic force along $x$ on the other test mass (TM2). All other degrees of freedom were also controlled, but the details are not relevant here. 

As the distance between the test masses was actively controlled, the out-of-the-loop differential disturbance force per unit mass acting on the test masses, $\Delta g_e$, could not be identified with the in-loop relative acceleration measurement $\Delta \ddot{x}$ \cite{armano:subfemtog}. This acceleration, which was estimated numerically \cite{derivative}, on the contrary had to be corrected for the known applied feedback forces per unit mass $g_c(t)$ \cite{armano:subfemtog}. 

In addition, acceleration data series were also corrected for the following effects.
\begin{enumerate}[label=(\roman*)]
    \item  Measured  inertial forces per unit mass  due to spacecraft rotation $g_i(t)$, which include the specific centrifugal force and the apparent angular acceleration \cite{PhysRevLett.120.061101}. These effects will not be relevant for LISA.
    \item The  forces per unit mass generated by the motion of the test masses through  static force gradients in the spacecraft, as LISA data  can also be corrected for those.  Such force acting on TM$i$ is well approximated by the linear model $-\omega_i^2 x_i$, as described in \cite{PhysRevLett.120.061101}. 
    \item The spurious pick up $g_\text{CT}(t)$ of spacecraft motion along different degrees of freedom, due to cross-talk \cite{PhysRevD.97.122002,hartig_geometric_2022}, which also includes the pick-up of the common mode motion of the test masses, described by a term $\delta_{x_1} \ddot{x}_1$.  
\end{enumerate}

Thus the  out-of-the-loop, differential force per unit mass data series used in the following analyses, can be written as:  
\begin{equation}
	\label{eq:deltag}
	\begin{split}
	    \Delta g(t)=& \Delta\ddot{x}_\OMS(t) +\omega_2^2 \Delta x_\OMS(t)+ (\omega_2^2-\omega_1^2) x_{1,\OMS}(t)\\& - g_c(t)-g_i(t) - g_\text{CT}(t).
	\end{split}
\end{equation}
Note that, in Eq.~\eqref{eq:deltag}, we have attached the suffix $\OMS$ to all coordinates, to indicate that these have been measured by the relevant interferometers, and not by the capacitive sensors. For these we will use the $\GRS$ suffix. Note also that all the $\omega$'s above, as well as $\delta_{x_1}$, have been measured in dedicated calibration experiments \cite{PhysRevD.97.122002}. In particular $\omega_2^2 \approx \SI{-4.5e-7}{s^{-2}}$ is negative, while the differential stiffness $(\omega_2^2- \omega_1^2)$ is roughly 20 times smaller and thus neglected in this discussion here.

 $\Delta g (t)$ in Eq.~\eqref{eq:deltag} is our best estimate for $\Delta g_e(t)$. However the series is corrupted by any disturbance  $n_\OMS(t)$, random noise or spurious signal, affecting  the differential interferometer readout $\Delta x_{\OMS}$. Such disturbance enters into $\Delta g(t)$ in Eq.~\eqref{eq:deltag}, both through $\Delta\ddot{x}_\OMS(t)$, and through $\omega_2^2 \Delta x_\OMS(t)$. Thus, the residual noise in $\Delta g$ can be evaluated:
 \begin{equation}
     \label{eq:deltag2}
     \Delta g = \Delta g_e(t)+\ddot{n}_{\OMS}(t)+\omega_2^2 n_\OMS(t).
 \end{equation}

For the sake of the following discussion, it is important to note the following.
\begin{enumerate}[label=(\roman*)]
    \item The suspension control loop  has significant gain only at low frequency, with a \SI{3}{\dB}  cutoff  at about \SI{4}{\milli\hertz}. Thus $ - g_c(t)$ becomes  a good representation of $\Delta g_e(t)+\ddot{n}_{\OMS}(t) +\omega_2^2 n_\OMS(t)$ for signals at \si{\milli\hertz} and below, while in the opposite limit, it is $\Delta\ddot{x}_\OMS$ that mostly contributes to $\Delta g(t)$. 
    \item $\Delta g$ measures the difference between  the forces  acting on the two test masses along $x$. Thus there is no way of discriminating the contributions of the individual test mass forces; for instance the differential acceleration could be entirely caused by force on one test mass or the other.
    \item $\Delta g >0$ corresponds to a force pushing the test masses one toward the other.
\end{enumerate}

\subsection{Data runs}
The data we consider here are made of uninterrupted \emph{runs} during which the test masses and the satellite were in steady control conditions, with no purposely applied stimulus of any nature\footnote{Only in one these, a two day long run, a small sinusoidal signal of less than \SI{100}{\femto\newton} amplitude was injected in the suspension control loop to calibrate its force authority. Its effect on $\Delta g$ was negligible.}.  We have analyzed data from the three different kind of runs listed below.
\begin{enumerate}
    \item \label{r1} The week-long runs, with physical conditions adjusted to reach the lowest noise level, which we used to estimate the quasi stationary noise PSD \cite{armano:subfemtog}. During the majority of these runs, the  temperature of the LTP was kept at about \SI{22}{\celsius}. However in two of these runs the temperature was lowered to about \SI{11}{\celsius}, to decrease the outgassing rate and hence  pressure and  Brownian noise. One of these is the best noise performance run of February 2017 \cite{PhysRevLett.120.061101}.
    \item \label{r2} Shorter runs, with operating conditions slightly different from those used to achieve the best noise performance, but still with low noise and quasistationary behavior. These also include runs during which the spacecraft control used an alternative set of microthrusters, based on colloidal propellants. In these runs the noise performance was slightly worse than in the low noise runs \cite{PhysRevD.98.102005}.  Runs in \ref{r1} and \ref{r2} include a total measurement time of $\SI{138.4}{\day}$. 
    \item \label{r3} A set of runs of lower stability, the origin  of which we describe as follows. In May 2017, in an attempt to further decrease  pressure, we lowered the temperature to about \SI{0}{\celsius}, a value outside the nominal operating range of the instrument. The instrument entered into a rather  unstable state, with a  rate of glitches so high as to make the estimate of the background noise meaningless below $\sim \SI{1}{\milli\hertz}$. When the temperature was raised again to \SI{11}{\celsius}, the instrument went back to its ordinary behavior. These runs lasted a total of \SI{11.9}{d}.
\end{enumerate}
In the rest of the paper we call the runs in \ref{r1} and \ref{r2}  \emph{ordinary runs}, and those in \ref{r3}, at about \SI{0}{\celsius}, \emph{cold runs}.

\section{\label{sec:gltchdetection} Glitch detection and parameter estimation}
Glitches were detected as localized signals in the $\Delta g(t)$ data series,  emerging from noise after some data pre-processing. They fall in two broad categories: impulse carrying glitches, and high frequency, low-impulse glitches. The detection method and the parameters we estimate are different for the two categories. We describe both in the following. 
\begin{figure*}[thpb]
    \vspace{3mm}
    \includegraphics[width=0.95\textwidth]{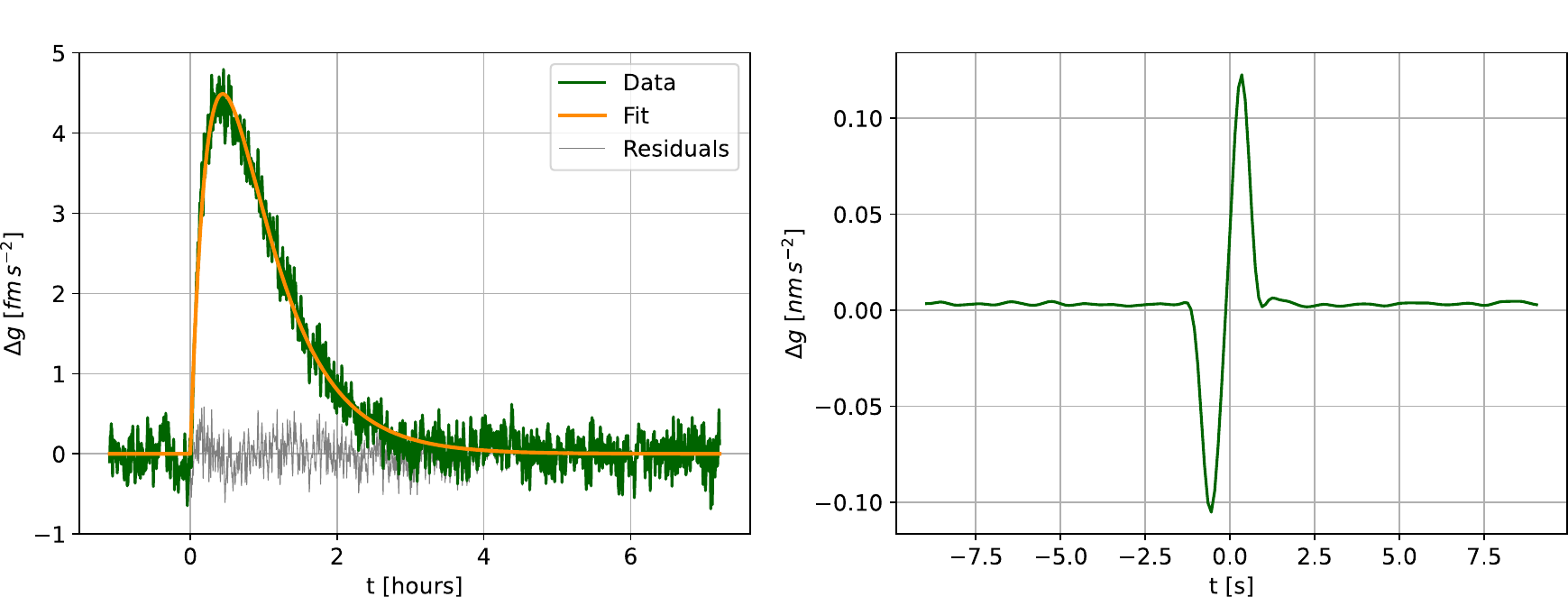}
    \caption{Left: example of an impulse carrying glitch. The picture shows: (green) the native data after low-pass filtering and background subtraction; (orange) the fitting template; (gray) the residual after subtraction of the template. Right: example of a fast, low impulse glitch. The figure shows data after the $\SI{0.5}{\hertz}$ low-pass filtering.} 
    \label{fig:glitch_eg}
 \end{figure*}
\subsection{\label{sec:imp}Impulse-carrying glitches}

The first category includes \numonesided signals (98 in ordinary runs, 334 in cold runs) that  carry  significant  total  impulse per unit mass $\Delta v=\int_0^\infty \Delta g(t)\,\mathrm{d}t$. 

Note that  $\Delta v$ is also the zero-frequency limit of Fourier transform of the $\Delta g$ signal. Thus these signals have  significant energy within the low frequency  band of the control loop, and hence show up also as features within the feedback force data series  $g_c(t)$.

We manually detected glitches of this kind by first filtering the data with a low-pass, finite impulse response filter, consisting of a $\SI{100}{s}$-long Blackman-Harris window with an effective roll-off frequency at $\SI{10}{\milli\hertz}$. 

The filter suppressed the intense high frequency noise coming from the double time derivative of the interferometer readout noise, and made the glitch fairly visible above the remaining noise within some continuous stretch of the $\Delta g(t)$ data series, that we call the glitch stretch. 

Subsequently, for each glitch, a polynomial was fitted to the data  in two $\SI{1000}{\second}$ long stretches, one immediately preceding the glitch stretch, and the other immediately following it. The polynomial was of first order for glitch stretches shorter than $\SI{1000}{\second}$ (see later for the definition of duration), and of second order for longer glitches. The best-fit polynomial was then subtracted from the data. The result of this procedure, for one of the longest glitches, is shown in Fig.~\ref{fig:glitch_eg}, left panel.   

Such background subtraction  was necessary to get rid of the long term  drift that affected all data,  mostly due to the gravitational signal from propellant tank depletion and   long term temperature variations. The use of a second order polynomial for the longest glitches was able to accommodate some drift rate variation over many hours duration.

To these pre-processed data, we fitted  a simple signal template  (described further down) in the time domain, properly low-passed with the same filter used for the data (see Fig.~\ref{fig:glitch_eg}). For ordinary runs, the main purpose of such fitting was to remove the glitch from the data. This  allowed us to estimate the PSD of the underlying  noise on the entire data series, thus reaching  the lowest attainable frequencies. To this aim, we subtracted the unfiltered version of the best fit signal $h(t)$ from the native $\Delta g(t)$ time series. 

\begin{figure}[htbp]
    \centering
    \includegraphics[width=\columnwidth]{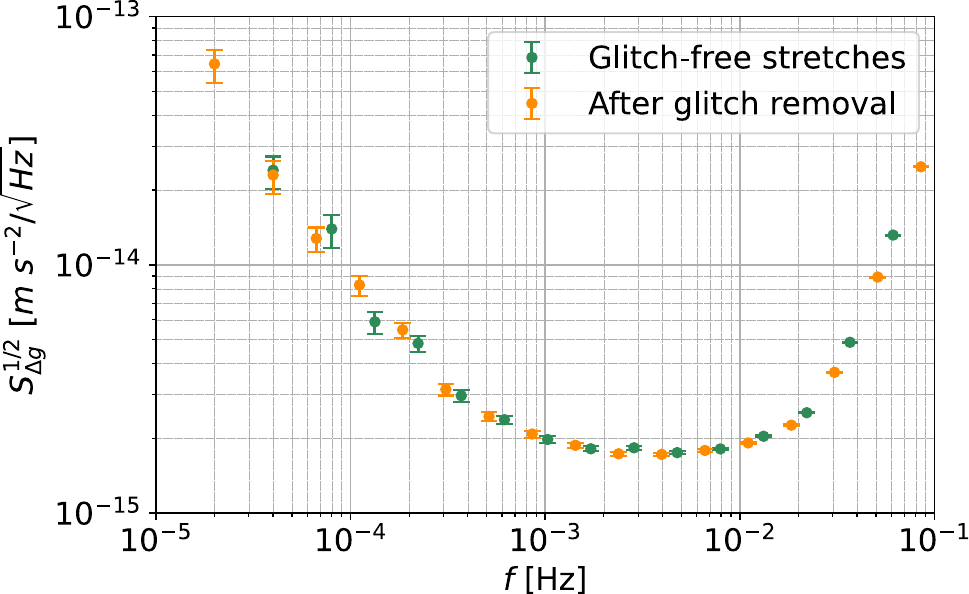}
    \caption{Amplitude spectral density (ASD) $S_{\Delta g}^{1/2}(f)$ of quasistationary noise, vs the frequency $f$, of the best noise run of February 2017. Green data points: ASD of glitch-free stretches only. Orange data points: ASD of residuals after glitch removal. Errors represent 68\% confidence intervals.}
    \label{fig:glitch_subtract_cutout}
\end{figure}

For ordinary runs, the procedure was indeed quite effective \cite{PhysRevLett.120.061101}, leaving a residual time series with the same PSD, within errors, as that of the series from which the glitch stretch was simply removed. This comparison is only possible down to the lowest frequency that could be attained with both time series (see Fig.~\ref{fig:glitch_subtract_cutout}).

For cold runs the data series were rather complex, with very short time intervals between glitches and quite a number of overlapping ones. Such complexity reduced the quality of the fitting and did not allow for glitch subtraction to any  level  useful for PSD estimation below \SI{1}{\milli\hertz}. Nevertheless the fit allowed estimating the glitch key parameters (see below).

Of \numonesided events, \numonesidedtwotau were fitted to the heuristic  template \cite{PhysRevLett.120.061101}:
\begin{equation}
    h_2(t) = \frac{\Delta v}{\tau_1-\tau_2} \left( e^{-t'/\tau_1} - e^{-t'/\tau_2} \right) \, \Theta(t'), \quad t'=t-t_0,
    \label{eq:onesidedshape_2tau}
\end{equation}
Here: $t_0$ is the glitch occurrence time; $\Theta(t)$ is the Heaviside step function; the decay time is set by the larger of the time constants  $\tau_1$ and $\tau_2$, while  the  rise time by the shorter of the two; finally,  $\Delta v$ is  the total transferred impulse per unit mass.

\newpage
For \numonesidedcontinuity of the \numonesided glitches, the  template in Eq.~\eqref{eq:onesidedshape_2tau} simplified to its limit\footnote{ Note that this template corresponds to the first-order shapelet considered in \cite{PhysRevD.105.042002}.} for $\tau_1\to \tau_2 \to \tau$: 
\begin{equation}
    \label{eq:onesidedshape}
    h_1(t) = \frac{\Delta v}{\tau^2} \,t'\, e^{-t'/\tau} \, \Theta(t'), \quad t'=t-t_0
\end{equation}

Two exceptional glitches, the longest ones and both in ordinary runs,  required a third exponential function
\begin{equation}
\begin{split}
    h_3(t) =
    \Delta v &\left(\frac{\tau_1 \,e^{-t'/\tau_1}}{(\tau_1-\tau_2)(\tau_1-\tau_3)}+\frac{\tau_2 \,e^{-t'/\tau_2}}{(\tau_2-\tau_3)(\tau_2-\tau_1)}+\right .\\&\left .+\frac{\tau_3 \,e^{-t'/\tau_3}}{(\tau_3-\tau_1)(\tau_3-\tau_2)}\right)\, \Theta(t'), \quad t'=t-t_0,
    \end{split}
    \label{eq:onesidedshape_3tau}
\end{equation}
to account for some fine structure of the glitch onset, and to allow for good quality glitch subtraction. 

For all templates above, and for  the  glitches fitted to them, $h(t)$ never crosses the level $h(t)=0$.
We found 3 glitches that  still carry a significant impulse, but, contrary to the others, do cross  $h(t)=0$ during their evolution, even though only once. These glitches are fitted to
\begin{equation}
    h_c(t) =h_2(t)+\tau_d \dot{h}_2(t) ,
    \label{eq:onesidedshape_zerocross}
\end{equation}
with $\tau_d$ a constant that can have any sign.

Note that all  templates above and their related glitches  leave  no net step in the $\Delta g$  data series after glitch subtraction, as the glitch signal  goes to 0 at  both its ends.


Given the nonoptimal nature of a time domain  fit in colored noise, we have not performed any dedicated goodness-of-fit test, besides the check on the PSD of residuals. In particular we have not made any systematic comparison among  different templates. Actually in some cases, for the shortest glitches of smaller amplitude, a fit with a filtered version of a Dirac delta or with a simple exponential gave comparable results.

In addition to allowing for glitch removal, the fitting also allowed us to estimate the values of  all template parameters and their errors, as described in Appendix~\ref{app:AEE}. From these parameter values, for the sake of further analysis, we also calculated an effective duration $\Delta$ for  the templates in Eqs.~\eqref{eq:onesidedshape_2tau} to~\eqref{eq:onesidedshape_3tau} as the time interval following $t_0$ in which there is  99\% of its signal energy, defined as $\int_0^{t} h^2(t') \, \mathrm{d}t'$. In the simplest case of Eq.~\eqref{eq:onesidedshape}, this corresponds to $\Delta \sim 4.20 \, \tau$. For the template in Eq.~\eqref{eq:onesidedshape_zerocross}  we define  the duration as that of $h_2(t)$ in that same equation.

\subsection{\label{sec:fast}Fast, low-impulse glitches}
A second population of glitches, which included  \numtwosidedOR glitches in ordinary runs and \numtwosidedCR in cold runs, is characterized by short duration and minimal total impulse. Specifically,
\begin{itemize}
    \item The overall duration of the glitch is compatible, within errors, with that of the convolution of the impulse response of the filter used to estimate the second time derivative, with that of the low-pass filter.
    \item They carry no significant impulse (per unit mass) $\Delta v_\text{glitch}=\int_0^\infty \Delta g(t)\mathrm{d}t$ (see Sec.~\ref{sec:dvtwosided}).  As a consequence, they show no detectable counterpart in  the feedback force time series $g_\text{c}(t)$.
    \item They are detected in the data after  low-pass filtering with a filter (Blackman-Harris, \SI{2}{s} long) with a roll-off frequency of $\simeq \SI{0.5}{\hertz}$, i.e., significantly higher than that used for the other category of glitches. Actually, the majority shows up with no filtering at all.
\end{itemize}

Of the \numtwosidedOR glitches of this kind observed in ordinary runs, 2 were well subtracted from the data  by fitting them to the filtered version of the second time derivative of a step in  the $\Delta x$ time series. As expected, given the high frequency nature and the lack of impulse, the subtraction had no detectable effect on the PSD. Given such a negligible impact on the data, the 2 remaining glitches that were discovered in a second search were neither fitted nor subtracted. 

This kind of glitch was not subtracted from cold runs data either, as the fitting turned out to be unfeasible in most of the cases. Thus, for the sake of the following analysis, for all glitches in this category we will only consider the  impulse $ \Delta v_\text{glitch}$  and the time of occurrence, defined as the time when $\Delta g(t)$ reaches the maximum of its absolute value.

\subsection{Other spurious signals in the data}
In addition to the glitches described above, we found a few  signals in $\Delta g$ caused by the impact of the spacecraft with micrometeoroids. The impact caused a well identified acceleration of the spacecraft \cite{Thorpe_2019}. As mentioned in Sec.~\ref{sec:dyn}, $\Delta g(t)$ has been corrected for the acceleration of the spacecraft. However, for the most energetic  events a residual signal was still found in the data due to calibration errors. Glitches of this kind are well understood and could have been suppressed by a better calibration. Thus we are not going to discuss them here any further.

We will also not discuss  spikes in the data, some of which were periodic, observed upon operation of particular devices and that could be reproducibly suppressed by turning their source off. 

\section{Glitch parameter statistics}
\label{sec:glitchparamstats}
We have performed a statistical analysis of the observed glitch parameters. In the following we give the results of such analysis, separately for the  different glitch categories.

\subsection{Impulse-carrying glitches}
In this section we present the main statistical features of the glitches that are fitted to the models of Eqs.~\eqref{eq:onesidedshape_2tau} to~\eqref{eq:onesidedshape_3tau}. We only briefly discuss at the end the properties of the few glitches fitted to Eq.~\eqref{eq:onesidedshape_zerocross}.
\subsubsection{Occurrence rate and waiting time}
\label{sec:poisson}
\begin{figure}[htbp]
    \centering
    \includegraphics[width=0.9\columnwidth]{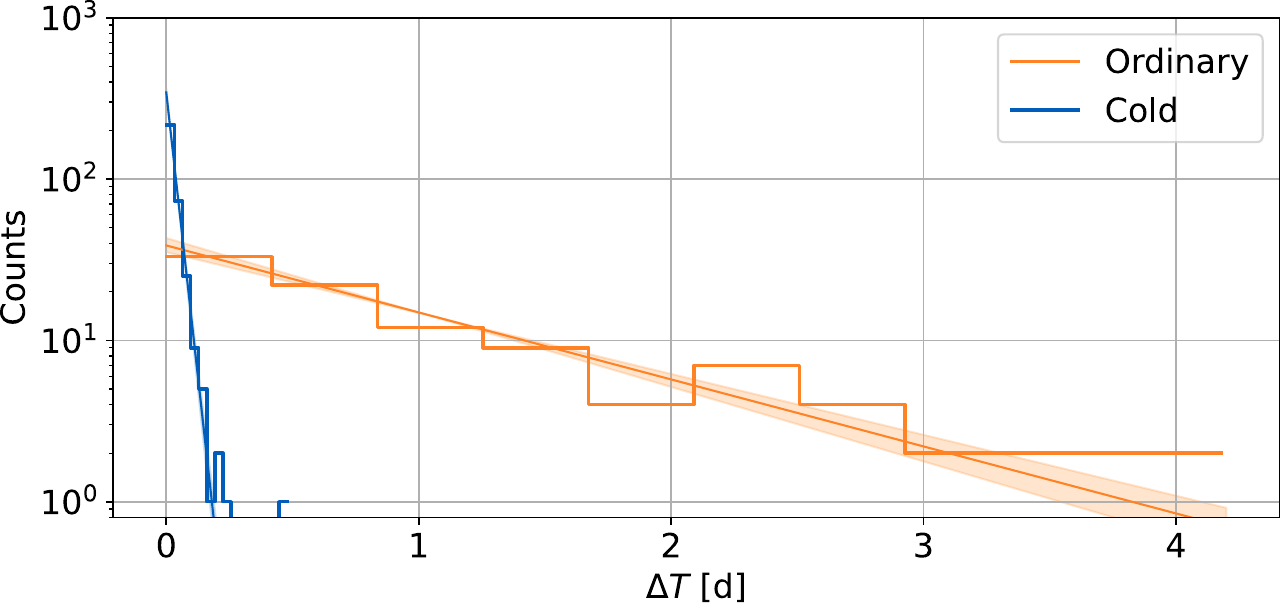}
    \caption{Histogram of the waiting time $\Delta T$ for ordinary runs (orange)  and cold runs (blue). Straight lines, and associated shadowed areas, represent, respectively, the Bayesian fit to an exponential distribution, and the corresponding 68\% confidence interval. For the cold runs, the fit is limited to data with $\Delta T \le  \SI{0.2}{\day}$. The rates from the Bayesian estimation for ordinary runs, and cold runs with $\Delta T \le  \SI{0.2}{\day}$, are, respectively, $\lambda=0.96_{-0.09}^{+0.11}\,\si{\day^{-1}}$ and $\lambda=32_{-2}^{+2}\,\si{\day^{-1}}$.}
    \label{fig:tauhistograms}
\end{figure}

Figure~\ref{fig:tauhistograms} shows the histograms of the waiting time $\Delta T$ between impulse-carrying  glitches for both ordinary and cold runs. $\Delta T$  is defined as either the time between two subsequent glitches, the vast majority of the samples, or the time between the starting moment of the run and the first glitch, relevant for very short runs with few glitches. 

For both ordinary and cold runs, the result of a Lilliefors test \cite{Lilliefors} is compatible with $\Delta T$ being exponentially distributed. Data for ordinary runs are well fitted to an exponential distribution with average rate $\lambda=0.96_{-0.09}^{+0.11}\,\si{\day^{-1}}$ (\footnote{This rate of $\lambda=0.96_{-0.09}^{+0.11}\,\si{\day^{-1}}$ is apparently slightly higher than that reported in our preliminary search \cite{PhysRevLett.120.061101} of $\lambda=(0.78\pm0.02)~\si{\day^{-1}}$. We have traced back this apparent discrepancy to the  smaller subset of  runs used in \cite{PhysRevLett.120.061101}, and to a mistake in reporting the error. The event rate estimated with the current Bayesian analysis, at 68\% confidence level on the same subset of runs, gives  $\lambda=0.75_{-0.09}^{+0.13}\,\si{\day^{-1}}$, which is compatible with the current estimation at $\simeq 1 \sigma$.}). Bayesian rate estimation is described in Appendix~\ref{app:RateLikelihood}.

Data for the cold runs, and for $\Delta T \le  \SI{0.2}{ \day}$, are also compatible with an exponential distribution with $\lambda=32_{-2}^{+2}\,\si{\day^{-1}}$. However the distribution shows a clear excess tail for longer times, amounting to some excess counts at longer times, see Fig.~{\ref{fig:tauhistograms}}, originating from data taken at different temperatures (see below).

Figure~\ref{fig:ordinaryrate} shows the time evolution of the rate for ordinary runs over the course of the mission, which is  consistent with a time independent value.

\begin{figure}[htbp!]
    \centering
    \includegraphics[width=0.87\columnwidth]{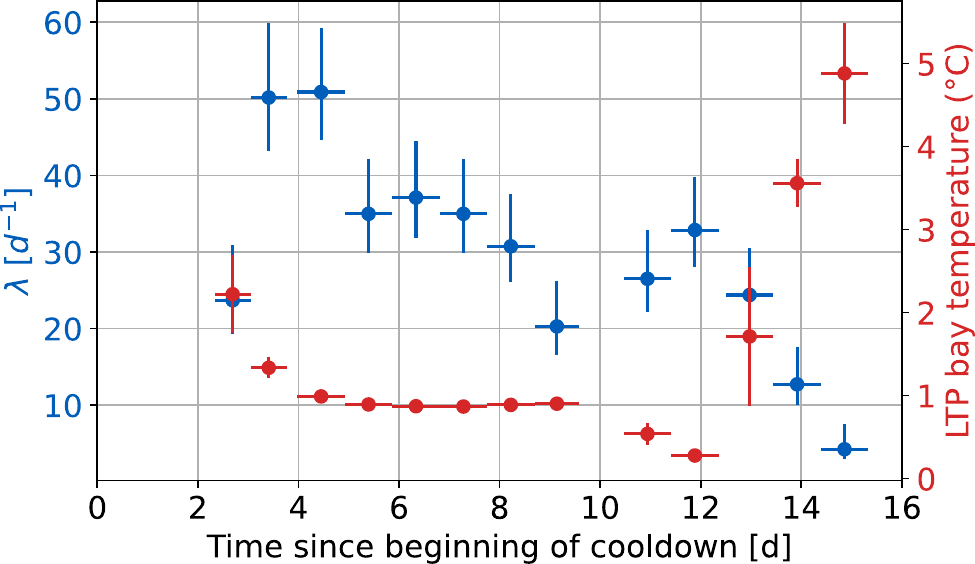}
     \caption{Glitch occurrence rate $\lambda$ (left scale), and LTP bay temperature (right scale), during cold runs, as a function of time since the beginning of cooldown. Small temperature changes before day 12 were the result of adjustments of  heaters settings aimed at stabilizing the system behavior. Reheating to ordinary conditions started at day 12.}
     \label{fig:coldseries}
\end{figure}

Figure~\ref{fig:coldseries} shows the evolution of the rate $\lambda$ and of  the system temperature $T_\text{LTP}$ during cold runs. The reported temperature is that of the bay that contained the LTP, as the readout electronics for all thermometers on the LTP itself saturated at about $\SI{8}{\celsius}$. 

Note that the rate variations of Fig.~\ref{fig:coldseries}  fully explain the few excess counts at longer waiting times in the cold runs histogram of Fig.~\ref{fig:tauhistograms}.

\begin{figure}[htbp]
    \centering
    \includegraphics[width=0.87\columnwidth]{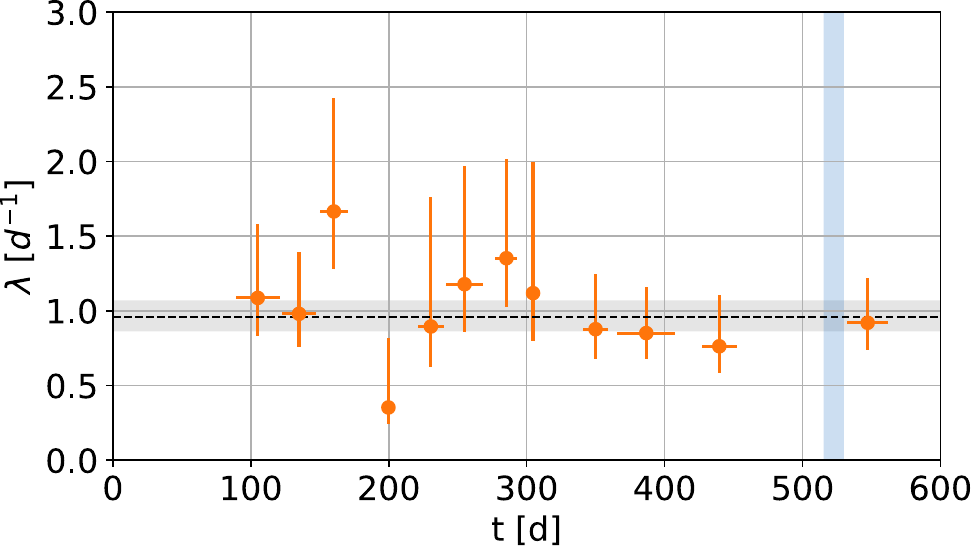}
    \caption{Glitch occurrence rate $\lambda$, during ordinary runs, vs time $t$ from launch, throughout the entire mission. Points are calculated by grouping glitches observed during runs the start times of which differ by less than a month. Vertical errors bars are Bayesian estimates assuming exponential distribution, and corresponds to 68.3\% (1$\sigma$) likelihood (see Appendix~\ref{app:RateLikelihood}). Horizontal error bars correspond to the total duration of  the considered epoch. The dashed line, and the associated gray shaded area, represent, respectively,  the mean rate and its error from the Bayesian estimate. The blue shaded area indicates the epoch of cold runs.}
    \label{fig:ordinaryrate}
\end{figure}
\vspace{-3mm}

\clearpage
\subsubsection{Impulse and duration}
\begin{figure*}[bthp!]
    \centering
    \includegraphics[width=\textwidth]{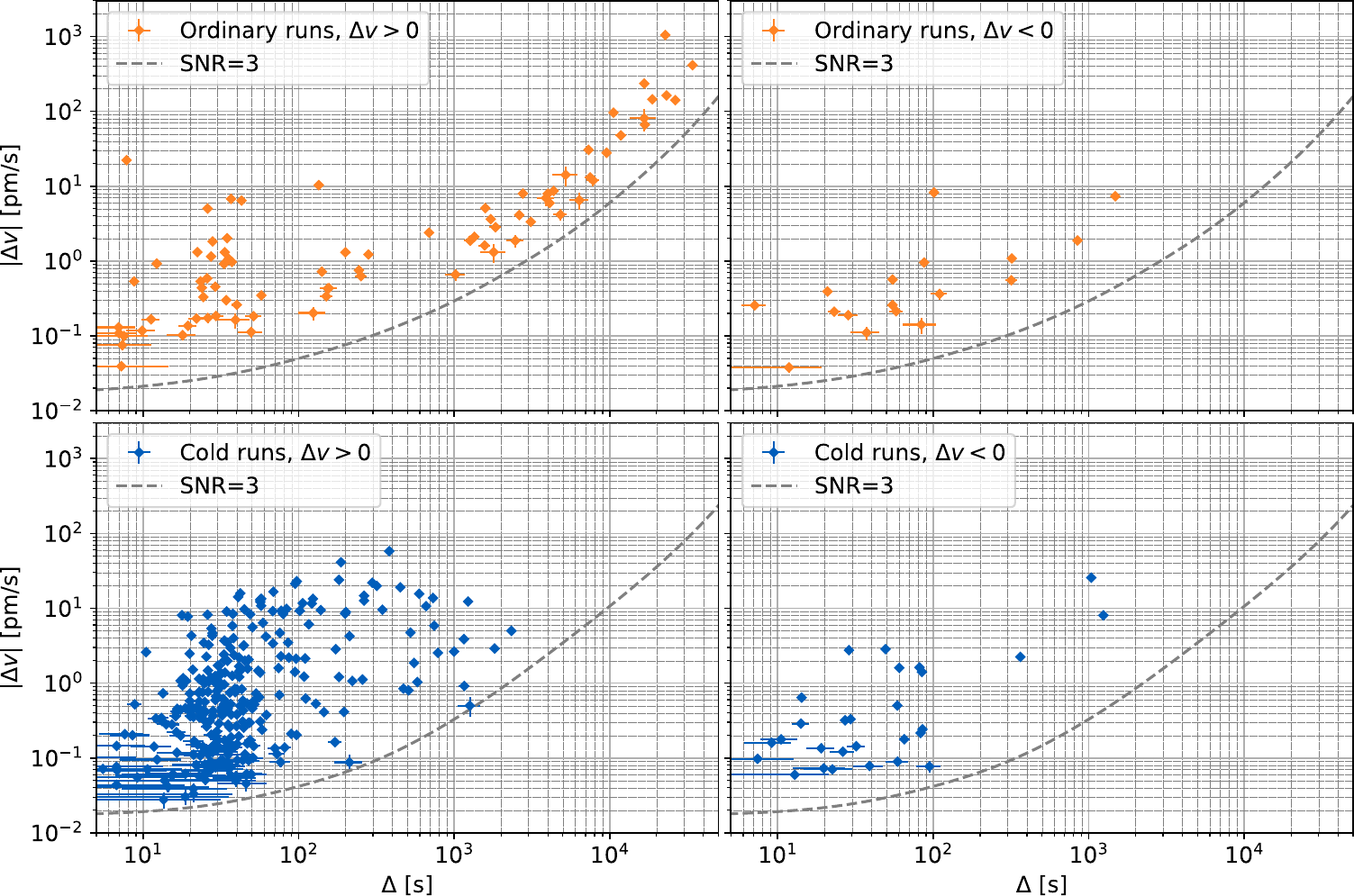}
    \caption{Absolute impulse per unit mass $\left|\Delta v\right|$, and  duration $\Delta$, of impulse carrying glitches fitted to templates in Eqs.~\eqref{eq:onesidedshape_2tau} to~\eqref{eq:onesidedshape_3tau}. 
    Upper left panel: the 81 positive impulse glitches observed during ordinary runs.
    Upper right panel: the 17 negative impulse glitches observed during ordinary runs. 
    Lower left panel: the 306 positive impulse glitches observed during cold runs. 
    Lower right panel: the 28 negative impulse glitches observed during cold runs. 
    For reference, the gray dashed line represents, for any given duration, the amplitude of a glitch of the kind in Eq.~\eqref{eq:onesidedshape} that would have $\text{SNR}=3$. In the two upper panels, SNR is calculated for the lowest noise ordinary run of February 2017. In the lower panels, the line is calculated for the sensitivity of the cold runs. For a detailed analysis of the glitch SNR, see Fig.~\ref{fig:SNRhis}.}
    \label{fig:Glitch_LPF_spSNR3}
\end{figure*}

Figure~\ref{fig:Glitch_LPF_spSNR3} shows the impulse and the duration of all glitches that are fitted to the templates in any of Eqs.~\eqref{eq:onesidedshape_2tau} to~\eqref{eq:onesidedshape_3tau}.
For reference the figure reports, for any given duration,  also  the amplitude of a glitch, well fitted to the template of Eq.~\eqref{eq:onesidedshape} that would have $\left|\Delta v\right|/\sigma_{\Delta v}\equiv \text{SNR}=3$. Here $\sigma_{\Delta v}$ is the error on $\Delta v$, and SNR stands for signal-to-noise ratio. For ordinary runs, the line refers to the sensitivity of the February 2017 run, and is then a lower limit for the other, less sensitive runs.
\vfill

\onecolumngrid
\vspace{12mm}

\newpage
\twocolumngrid

\begin{figure}[htbp]
    \centering
    \includegraphics[width=\columnwidth]{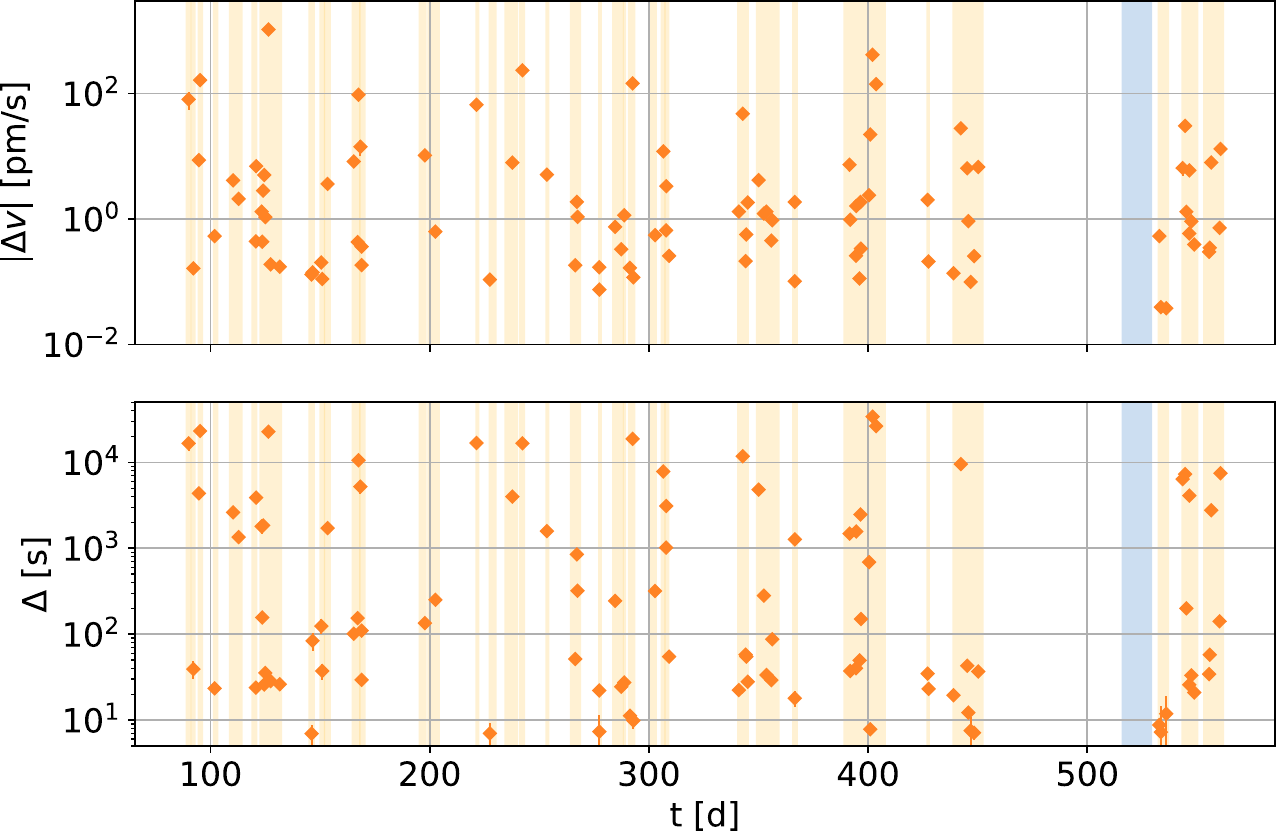}
    \caption{Absolute impulse per unit mass $\left|\Delta v\right|$ (top),  and duration $\Delta$ (bottom) as a function of time from launch for the glitches of   Fig.~\ref{fig:Glitch_LPF_spSNR3}. Note that the apparent clustering corresponds to the different measurement runs. The blue shaded area corresponds to the epoch of cold runs.}
    \label{fig:paramevol}
\end{figure}

In addition, Fig.~\ref{fig:paramevol} gives separately the time evolution of $\left|\Delta v\right|$ and $\Delta$ for the  glitches in ordinary runs.

\begin{figure}[htbp]
    \centering
    \includegraphics[width=\columnwidth]{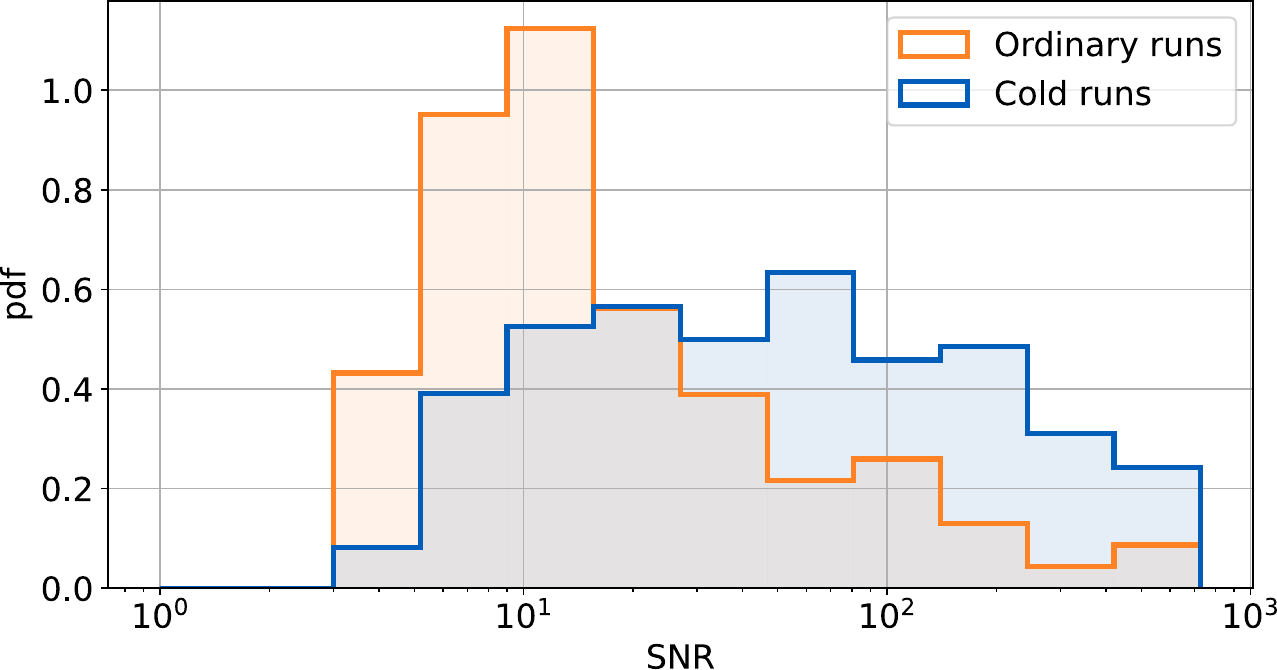}
    \caption{Histogram of the SNR for all glitches of Fig.~\ref{fig:Glitch_LPF_spSNR3}. The lowest observed value is $\text{SNR}\sim3$. The probability density refers to the logarithm of SNR.}
    \label{fig:SNRhis}
\end{figure}

Figure~\ref{fig:SNRhis} shows the histogram of their SNR, with its evident lower bound at $\text{SNR}\sim3$.

\begin{figure}[htbp]
    \centering
    \includegraphics[width=\columnwidth]{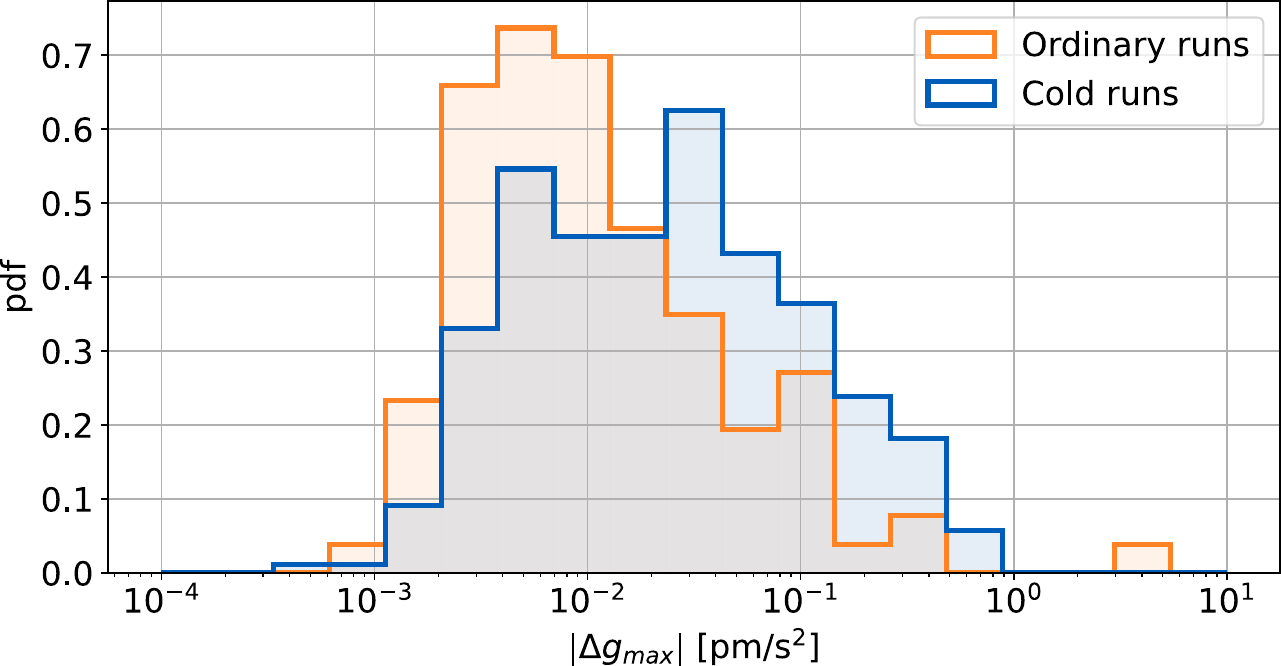}
    \caption{Histogram of peak amplitude $|\dgmax|$ for all glitches of Fig.~\ref{fig:Glitch_LPF_spSNR3}. The probability density is for the logarithm of the amplitude.}
    \label{fig:peakhis}
\end{figure}

For the sake of further discussion, we also report in Fig.~\ref{fig:peakhis} the histogram of the absolute peak value $\left|\dgmax\right|$ for the  glitches in Fig.~\ref{fig:Glitch_LPF_spSNR3}.

\subsubsection{Zero-crossing glitches}
The preceding figures do not include the 3 glitches that are fitted to the template in Eq.~\eqref{eq:onesidedshape_zerocross}. We summarize their properties in Table~\ref{tb:zerocrossing}.

\renewcommand{\arraystretch}{1.2}
\begin{table}[htbp!]
\caption{\label{tb:zerocrossing}
Observed glitches corresponding to the template in Eq.~\eqref{eq:onesidedshape_zerocross}. Values above the double horizontal line refer to ordinary runs. Values below that same line, to cold runs.}
\begin{ruledtabular}
\begin{tabular}{cccc}
$\Delta v$ [pm/s] & $\Delta$ [s] & $\tau_d$ [s] \\
\hline
\num{0.10\pm0.01}  & \num{35\pm8} & \num{-37\pm9}\\
\num{0.81\pm0.03}\footnotemark[1]  & \num{111\pm3} & \num{72\pm4}\\
\hline\hline
\num{0.42\pm0.02}  & \num{143\pm50} & \num{-29\pm22}\\
\end{tabular}
\end{ruledtabular}
\footnotetext[1]{Corresponds to an event in $\ddot{x}_{1,\OMS}(t)$, see Sec.~\ref{sec:IFOcorr}.}
\end{table}

\subsection{Fast, low-impulse glitches}
\label{sec:dvtwosided}
\begin{figure}[htbp]
    \centering
    \includegraphics[width=\columnwidth]{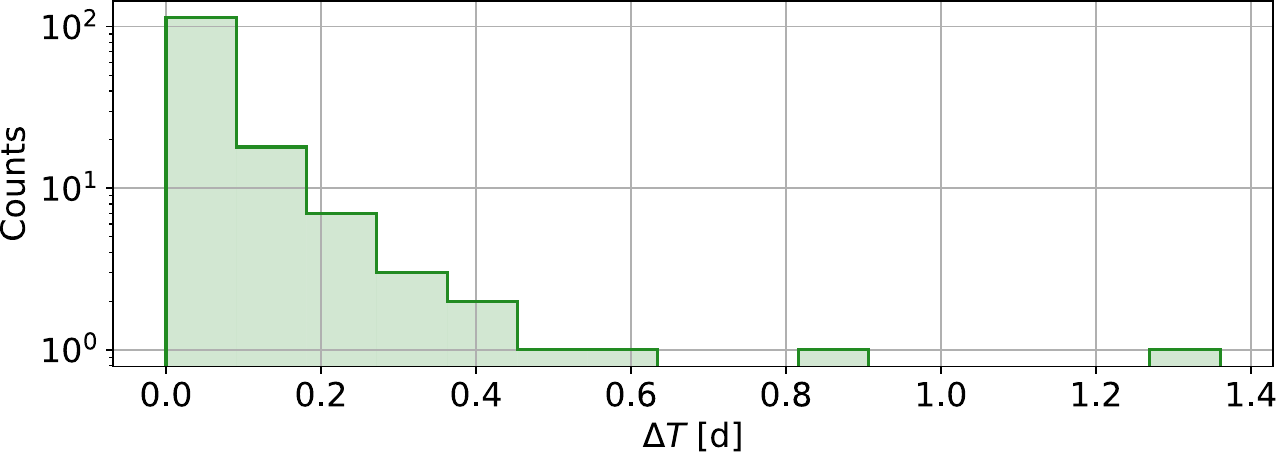}\\
    \includegraphics[width=\columnwidth]{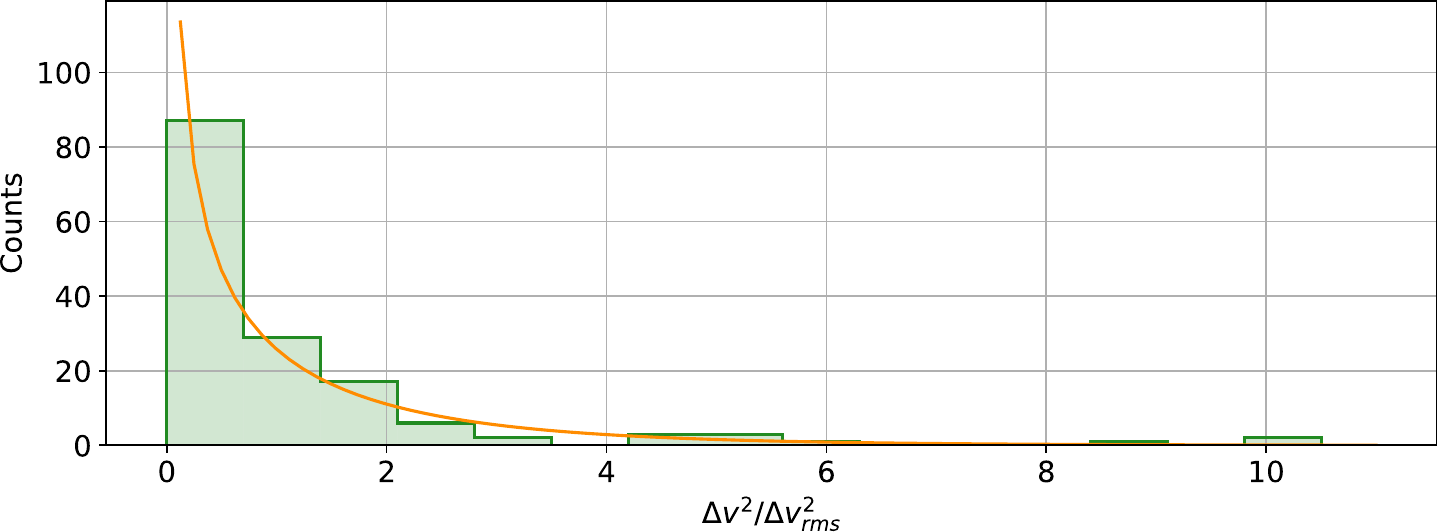}
    \caption{(upper) Histogram of the waiting time $\Delta T$ for fast, low impulse glitches in cold runs.\\ (lower) Histogram of $\left(\Delta v_\text{glitch}^2/\Delta v_\text{rms}^2\right)$ for low impulse glitches. The orange line is the properly normalized distribution for a chi-square with one degree of freedom. }
    \label{fig:deltaThis2sided}
\end{figure}

As glitches of this category were rare in ordinary runs, we limit our analysis to cold runs. Figure~\ref{fig:deltaThis2sided} in the upper panel shows the histogram of the waiting time. 

A Lilliefors test \cite{Lilliefors} for the  exponential distribution on such histogram fails, indicating some departure from Poisson statistics. Some subsets of the glitches show indeed a marked clustering, especially near the beginning of the reheating procedure.

We must note though that the mean rate was  different at different times  during the cold runs, while the test assumes a single  global distribution. A test on subsets of the data show  that some subset is still compatible with an exponential distribution.

Figure~\ref{fig:deltaThis2sided} in the lower panel reports the histogram of the impulse  $\Delta v_\text{glitch}$, estimated by numerically integrating the short data stretch containing the glitch,  normalized to its estimated error  $\Delta v_\text{rms}$. We  derive  $ \Delta v_\text{rms} $, by performing the same numerical integration used to estimate $\Delta v_\text{glitch}$, on random selected, glitch-free stretches of data, of same length and within the same run.

The histogram is quantitatively compatible ($p=0.23$) with a chi-square distribution with one degree of freedom, that is, with the hypothesis that $\Delta v_\text{glitch}$ is normally distributed with zero mean and standard deviation equal to $\Delta v_\text{rms}$.

\section{Joint analysis with other time series}
\label{sec:jointanalysis}
In an attempt to understand the nature of glitches we have analyzed some other  data series that have been  measured  synchronously with $\Delta g $ throughout the mission.

\subsection{\label{sec:grsoms} Discriminating between force and readout effect using the GRS \texorpdfstring{$\Delta x$}{Dx} sensor.}

The motion of both test masses along $x$ and relative to the spacecraft has been measured at all times by the capacitive sensor of the GRS. From their measured coordinates $x_{1,\GRS}(t)$ and $x_{2,\GRS}(t)$, we have formed a measurement of their relative displacement, independent of $\Delta x_\OMS(t)$:
\begin{equation}
    \Delta x_\GRS(t)=x_{2,\GRS}(t)-x_{1,\GRS}(t)+n_\GRS(t)
\end{equation}
with $n_\GRS(t)$ the measurement noise.

The difference between these two measurements  only contains the difference between the noise terms:
\begin{equation}
\Delta x_\OMS(t)-\Delta x_\GRS(t)=n_\OMS(t)-n_\GRS(t),
\end{equation}
and  would immediately reveal a spurious signal within $n_\OMS(t)$, if such signal were large enough to be detected against the relatively noisy GRS data.

We have used the $\Delta x_\OMS(t)-\Delta x_\GRS(t)$ data series  to discriminate glitches that may have been caused by such signals within $n_\OMS(t)$, from  those due to the true force $\Delta g_e(t)$.

Specifically, we have used the information that the glitch signal  $\ddot{n}_\OMS(t)+\omega_2^2 n_\OMS(t)$, would follow one of the templates in Eqs.~\eqref{eq:onesidedshape_2tau} to~\eqref{eq:onesidedshape_3tau}. 
If we define $h(s)$ to be the Laplace transform of such a template in $\Delta g$, then the Laplace transform of the relevant associated glitch in $n_\OMS(t)$ would be given by $h(s) / \left(s^2 + \omega_2^2\right)$. 

As $h(s)$ is in all cases a rational function of $s$, then $n_\OMS(t)$ would carry a diverging term  $\propto e^{+\sqrt{-\omega_2^2}t}\simeq e^{+t/\SI{1.5}{ks}}$, which after a few thousand seconds would dominate the data. 

\begin{figure}[htbp]
    \centering
    \includegraphics[width=\columnwidth]{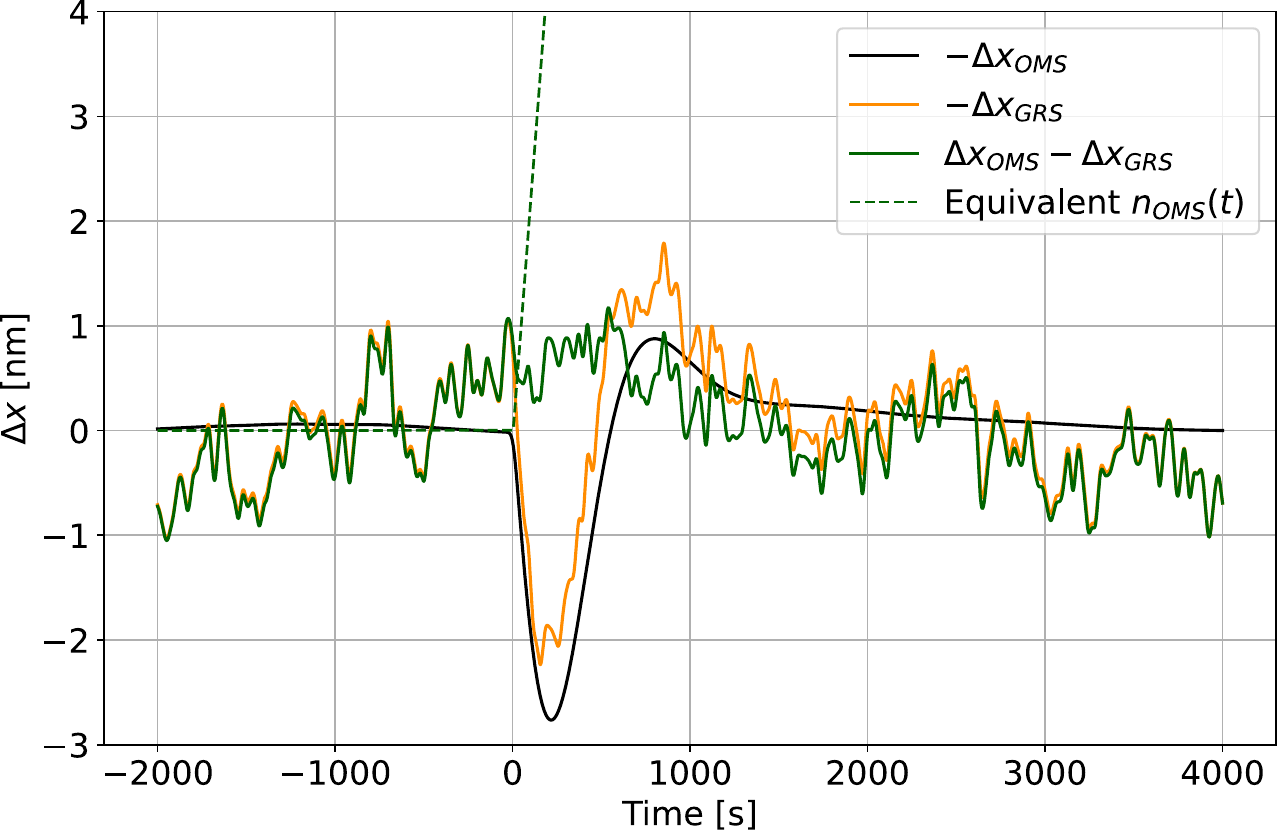}
    \caption{$-\Delta x_\OMS(t)$ (black), $-\Delta x_\GRS(t)$ (orange), and their difference (green) for an impulse-carrying glitch ($\Delta v=\SI{22.1}{pm/s}, \Delta=\SI{7.82}{s}$). The negative signs on the first two data series have been used to show that the residuals in the difference are dominated by the large noise in $\Delta x_\GRS(t)$. Also shown (dashed line) is the  signal one would observe in $\Delta x_\OMS(t)-\Delta x_\GRS(t)$, if the glitch were due to a spurious signal in $n_\OMS(t)$.}
    \label{fig:dgGRSOMS61}
\end{figure}

We illustrate the concept, for one of the glitches, in  Fig.~\ref{fig:dgGRSOMS61}. The glitch is clearly visible in both  $\Delta x_\GRS(t)$ and $\Delta x_\OMS(t)$, while it disappears in  their difference. In the figure we also show the inverse Laplace transform of $h(s)/\left(s^2+\omega_2^2\right)$, i.e. the signal one would observe in $\Delta x_\OMS(t)-\Delta x_\GRS(t)$, if the glitch were due to a spurious signal in $n_\OMS(t)$. For the  glitch in question, the picture clearly shows that this source for the glitch is ruled out. 

We have calculated such hypothetical signal in $n_\OMS(t)$, that is, the inverse Laplace transform of $h(s)/\left(s^2+\omega_2^2\right)$ for all glitches, both in the ordinary and cold runs. We  have found  that such signals would have been clearly visible in the data  if a sufficiently long observation time were available after the glitch occurrence time (up to $10^4~\si{s}$ for the weakest glitches). On the contrary, we have found none in the actual data. Only for a  few events in the cold runs we could not reach any conclusion  due to lack of sufficient observational data after the glitch.

Thus impulse-carrying glitches are due to true forces acting on the test masses, and are not artifact due to motion readout.

Low momentum, high frequency glitches consist of a feature in $\Delta x_\OMS$, and are not visible within the feedback force time series. However, they could still be either a spike in true acceleration, or just a feature in the interferometer output. 

As already mentioned, glitches of this kind consist of  a step, or of a few points outlier  in $\Delta x$, or some variation of those. In the case of just a feature in the interferometer, $n_\OMS(t)-n_\GRS(t)$ would then contain a similar feature (for these fast signals, we are neglecting  the term $\omega_2^2 n_\OMS$). Unfortunately the largest steps are a few tens of pm high, while the resolution on step detection in the $\Delta x_\OMS(t)-\Delta x_\GRS(t)$ is not better than $\sim \SI{1}{nm}$. Similar limitations hold for the detection of outliers. Thus we were not able to discriminate between true force and interferometer readout for this category of glitches.

\subsection{\label{sec:torque}Associated differential torque }

In close analogy with what we did with $\Delta g(t)$, we measured the differential out-of-the-loop torque per unit moment of inertia on the test masses, both around $y$ and around $z$. We have built these quantities by subtracting the control torques per unit moment of inertia from the measured differential angular accelerations of the test masses. For instance such differential torque for the $z$-axis, $\Delta\gamma_\phi$, is defined as
\begin{equation}
\label{eq:delta_gamma}
   \Delta\gamma_\phi = \ddot{\phi}_2-\ddot{\phi}_1+\left(N_{\phi_1}-N_{\phi_2}\right)/I_{zz},
\end{equation}
where $N$ are the commanded torques, $I_{zz}$ is the common moment of inertia around $z$, and the subscripts indicate the test mass. The subscript $\eta$ indicates the rotation about the $y$-axis.

As for the case of $\Delta g$, angular accelerations are corrected for some small gradient effect. More important, as the rotational motion of the spacecraft is rather intense, and is a common mode for both test masses, torques and angular rotations have been  recalibrated  to maximize the  rejection of such a large common mode disturbance. 

For each  of the impulse carrying glitches within the $\Delta g(t)$ time series,  we have fitted both the $\Delta \gamma_\phi(t)$ and the $\Delta \gamma_\eta(t)$ time series to exactly the same template in Eq.~\eqref{eq:onesidedshape_2tau} and \eqref{eq:onesidedshape}, with same time of occurrence and same time parameters, leaving the amplitude $\Delta \Omega_{\phi,(\eta)}$, an effective increase in angular velocity, as its sole fitting parameter. In addition, we have also included in the  fitting model  a parabolic background.

\begin{figure}[htbp]
    \centering
    \includegraphics[width=\columnwidth]{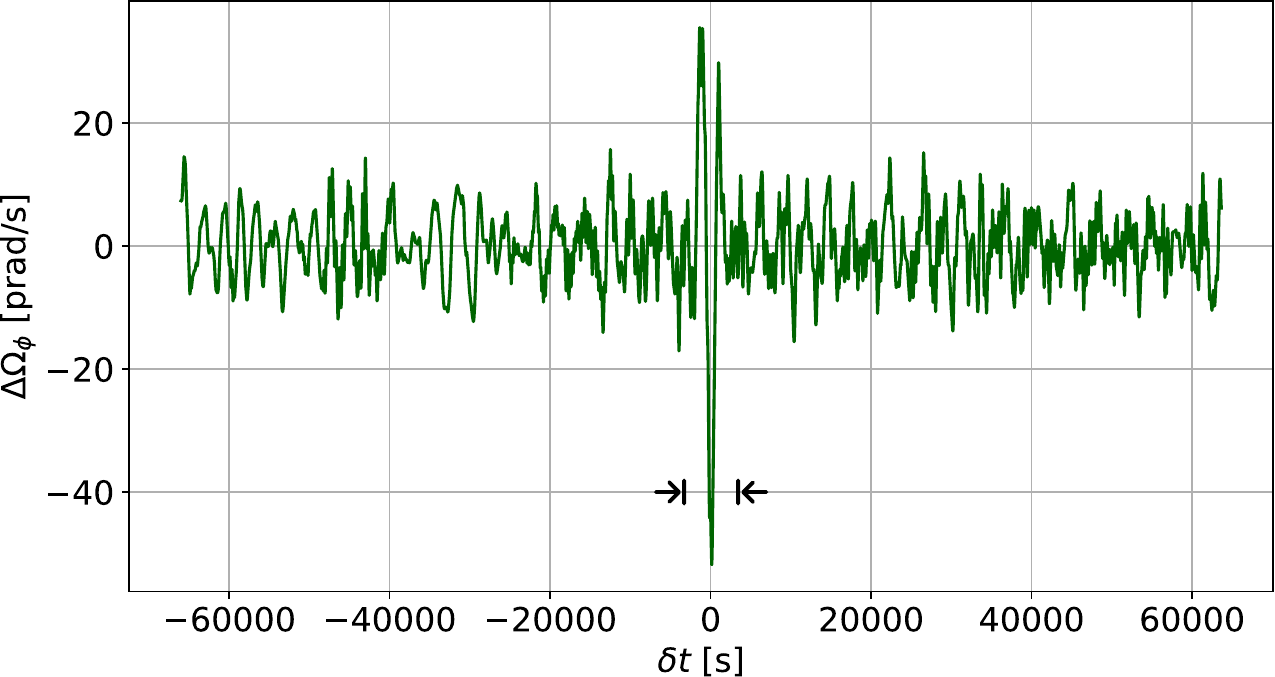}
    \caption{An example of the $\Delta \Omega_{\phi}(\delta t)$ series for one of the very few glitches with significant associated torque. The value in $\delta t=0$ is  the  result of  the fit for  native data. The arrows indicate the interval excluded for the sake of error evaluation.}
    \label{fig:omega}
\end{figure}

We have estimated the uncertainty on $\Delta \Omega_{\phi,(\eta)}$ by repeating the above fitting procedure  over the sliding stretch   $\Delta \gamma_{\phi,(\eta)}(t_1+\delta t,t_2+\delta t)$, with $t_1$ and $t_2$ the time bounds of the actual stretch that contains the glitch, and $\delta t$ a sliding delay. 
We have thus  generated  a series $\Delta \Omega_{\phi,(\eta)}(\delta t)$, with $|\delta t| \le 20(t_2-t_1)$, which  shows, at $\delta t=0$, a marked peak  above the background jitter, whenever the torque is significant (see Fig.~\ref{fig:omega}). 

As the $\Delta \Omega_{\phi,(\eta)}(\delta t)$ time series has an intrinsic autocorrelation over a scale $\delta t \simeq t_2-t_1$, it contains in practice only $\simeq 40$ independent data points. Thus its root mean square cannot be used,  as is, for error estimation. We then use, more cautiously, as an estimate for the error, the peak absolute value $\delta\Omega_{\phi,(\eta)}$ of the series, calculated on the  data  outside the central stretch $\left| \delta t\right| \le t_2-t_1 $. More specifically, if the central peak does not exceed, in absolute value, $\delta\Omega_{\phi,(\eta)}$, we take $\left|\Delta \Omega_{\phi,(\eta)}\right|\le\delta\Omega_{\phi,(\eta)}$.

If, on the contrary, the peak exceeds in absolute  value $\delta\Omega_{\phi,(\eta)}$, we take this as the error. It must be noted though that, for Gaussian statistics, the maximum absolute value among 40 independent samples falls in the interval $\left(2.4_{-0.4}^{+0.5}\right) \sigma$. Thus the confidence interval associated with such an error is greater than $ 95\% $.

A practical way to report  the results of this analysis, is to introduce an effective \emph{lever arm} defined as
\begin{equation}
    \label{eq:arm}
    r_{\phi,(\eta)}=\frac{I_{zz,(yy)}}{m}\frac{\Delta\gamma_{\phi,(\eta)}}{\Delta g}=\frac{I_{zz,(yy)}}{m}\frac{\Delta \Omega_{\phi,(\eta)}}{\Delta v} 
\end{equation}

For a single force, applied normal to one of the $x$ or $y$ faces of either test mass, $r_\phi$ would be the distance between the force application point and the center of the face. A similar interpretation holds for $r_\eta$. Note that for a real point-like force, given the size of the test mass, the maximum value for both $\left| r_\phi\right|$ and $\left| r_\eta\right|$ would be \SI{23}{mm}, while there is no upper limit for a distribution of forces. 

A particular interesting case, further discussed in Sec.~\ref{sec:disc}, is that of a force resulting from a voltage difference between the test mass and one of the electrodes facing its $x$-face. These electrodes were used for the control loop on TM2, but also for the angular control of both test masses. Such a voltage difference would have produced a force with a lever arm  $|r_\phi|\sim\SI{11}{mm}$  \cite{upperlimit}.

\begin{figure*}[htbp]
    \centering
    \includegraphics[width=\textwidth]{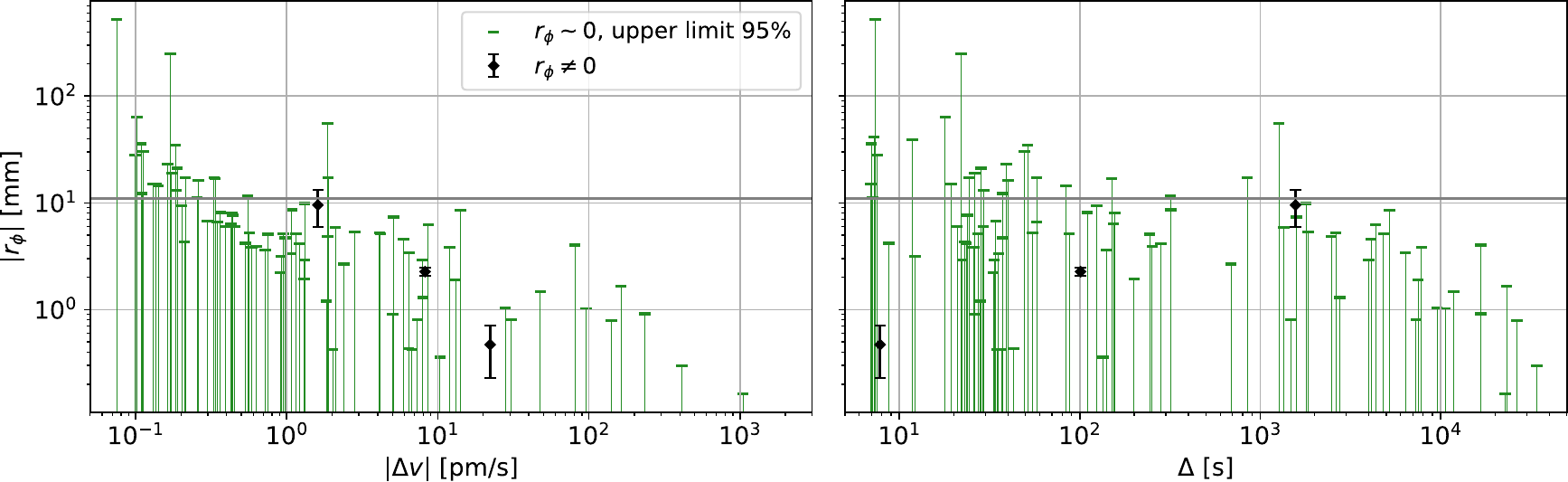}
    \caption{Absolute lever arm $\left|r_\phi\right|$, as defined in Eq.~\eqref{eq:arm}, as a function of  absolute impulse $\left|\Delta v\right|$ (left) and duration $\Delta$ (right), for impulse-carrying glitches of ordinary runs. The black data points represent the glitches for which the lever arm has been found to be significantly different from zero. The green horizontal segments represent the upper bound for $\left|r_\phi\right|$  for glitches for which $\left|\Delta \Omega_{\phi}\right|\le\delta\Omega_{\phi}$. For 9 of the 98 impulse-carrying glitches of the ordinary runs, we were not able to perform the analysis on the $\Delta\gamma_\phi(t)$ data series. The gray horizontal line refers to the \SI{11}{mm} ``electrical'' reference, as described in text.}
    \label{fig:arm}
\end{figure*}

The results of the analysis for $r_\phi$ in ordinary runs are shown in Fig.~\ref{fig:arm}. A list of the glitches with any of the lever arms significantly different from zero is reported in Table~\ref{tb:eventtorque}.

Figure~\ref{fig:arm} shows that, as expected from Eq.~\eqref{eq:arm}, the error on $r_\phi$ decays approximately as $\left|\Delta v\right|^{-1}$, decreasing below about \SI{1}{cm} for $\left|\Delta v\right| \gtrsim \SI{2}{pm/s}$.
For only 3 of the 98 impulse-carrying glitches in ordinary runs we detect an effective armlength that is significantly different from zero. In a slight majority (62\%) of the cases, where we find an armlength compatible with zero, we have sufficient resolution to exclude the hypothesis of an electrode origin to the glitch. Specifically, 62\% of the glitches in ordinary runs are both compatible with $r_\phi=0$ and have ``upper limit'' uncertainty of less than \SI{7}{mm} (or 2/3 of the characteristic ``electrode arm'' of \SI{11}{mm}). If any of these relatively high SNR glitches had had an \SI{11}{mm} effective armlength, they would have been detected with peaks more than 50\% above our background in Fig.~\ref{fig:omega}.

This fraction decreases to  50\% for  glitches with  $\Delta \le \SI{1}{\minute}$,  while it increases to 91\% for glitches with duration $\Delta \ge \SI{1.6}{ks}$, a duration threshold that will be discussed in Sec.~\ref{sec:disc}.

$\Delta\gamma_{\eta}(t)$ is significantly more noisy than $\Delta\gamma_{\phi}(t)$. This gives larger errors on $r_{\eta}$ than on $r_{\phi}$, and makes less likely to find lever arms significantly different from zero. For instance, just for only about 32\% of the glitches the error upper bound is smaller than $(2/3)\,\SI{11}{mm}$. We give  this figure here just for the sake of comparison with the case of $r_{\phi}$, but  $r_\eta=\SI{11}{mm}$ has no special significance.

The complexity of data during cold runs applies also to the $\Delta\gamma_{\phi,(\eta)}(t)$  series that contain multiple fast events and other non stationary features. This makes the results of our search procedure more noisy, reducing, for instance, the fraction of glitches, for which the upper error bound to $r_\phi$ is less than $(2/3)\,\SI{11}{mm}$, to about 45\%, and to 39\% if $\Delta \le \SI{1}{\minute}$.

Moreover, we note that, of the 3 zero-crossing glitches transferring impulse in Table~\ref{tb:zerocrossing}, none of them is associated to a measurable torque, measured with the same analysis as in this section. The error on $r_\phi$ associated to the stronger ones (one in ordinary runs and one in cold runs), is lower than $(2/3)\,\SI{11}{mm}$. 

As a final note, in addition to the glitches listed in Table~\ref{tb:eventtorque}, our procedure also finds 3 glitches in the cold runs for which the peak in $\Delta \Omega_{\phi,(\eta)}(t)$ is significantly displaced from the true time of occurrence. We believe that these   events are due the  accidental coincidence between a force glitch and an unrelated feature in the torque data. Thus we have not included them in the table.

\renewcommand{\arraystretch}{1.2}
\begin{table}[htbp!]
\caption{\label{tb:eventtorque}
Glitches with any of the lever arms significantly different from zero. Figures above the horizontal line refer to ordinary runs, figures below the same line to cold runs. Errors correspond to confidence $> 95\%$ (see text).}
\begin{ruledtabular}
\begin{tabular}{cccc}
$\Delta v$ [pm/s] & $\Delta$ [s] & $r_\phi$ [mm] & $r_\eta$ [mm] \\
\hline
$-8.23 \pm 0.03$ & $101.1 \pm 0.4$  & $-2.3 \pm 0.2$ & \dots \\
$1.6 \pm 0.2$    & $1560 \pm 140$   & $-9.5 \pm 3.5$ & \dots \\
$22.22 \pm 0.01$ & $7.82 \pm 0.01$& $ 0.5 \pm 0.2$ & $-1.2 \pm 0.4$ \\
\hline
$-25.68 \pm 0.02$  & $2100 \pm 5$   & $-1.2 \pm 0.7$ & \dots \\
\end{tabular}
\end{ruledtabular}
\end{table}

\subsection{Other interferometer channels}
\label{sec:IFOcorr}
\subsubsection{Low-impulse glitches}
We find  a significant number of coincidences between the low impulse glitches and the various interferometer channels, $x_{1,\OMS}$, frequency, and reference (see Sec.~\ref{sec:exp}).

\vspace{\baselineskip}
\noindent\textit{Reference/frequency channels.}---Out of 152 low impulse glitches detected in cold runs, some correspond to events in the reference or frequency channels, respectively 55 and 4. In the latest part of the cold runs, tens of glitches of this kind appear with non-Poissonian arrival time after the heaters were turned on to bring the temperature back to standard operating conditions. Those events showed a rapid deviation in the reference channel, with a characteristic shape lasting some tens of seconds due to the internal control loops of the optical interferometer system \cite{hechenblaikner_digital_IFO_2011}. The shape in either reference or frequency channels is then different from that of $\Delta g$. 

\vspace{\baselineskip}
\noindent\textit{$x_1$ channel.}---Some of the low impulse glitches belonging to cold runs were found to coincide with fast events in the $x_{1,\OMS}$ interferometer, detected as fast features in $\ddot{x}_{1,\OMS}(t)$. Among the 152 two-sided glitches detected in cold runs, 28 showed counterpart in $\ddot{x}_{1,\OMS}(t)$. 

The total number of glitches showing a coincidence in the interferometric channels is 81/152, given that some of them show multiple coincidences. However due to the complexity of the data of all time series, some coincidences may have gone undetected.

Out of the 4 low impulse events detected in ordinary runs, 2 showed coincidence to events in the interferometer time series, one in reference and the other one in $x_{1,\OMS}$. 
\subsubsection{Impulse-carrying glitches}
We find only one coincidence between an  impulse-carrying glitch and the other   interferometer channels. 
This is the coincidence between  the  zero-crossing glitch on the second line of Table~\ref{tb:zerocrossing} and a spike in the $\ddot{x}_{1,\OMS}(t)$ time series. The spike had no detectable counterpart in the time series of the force commanded onto the spacecraft by the drag-free control loop, as expected for such a fast feature.

In analogy with the case of the differential measurement, a spike like this might be either due to a spike in $g_1-G$, with $g_1$ the force per unit mass on TM1 along $x$, and $G$ that on the spacecraft, or to a feature in the $x_{1,\OMS}$ interferometer readout. When interpreted as a force, its total impulse, $\Delta v\simeq  \SI{0.8}{\nano \meter/ \second }$, is much larger than the corresponding impulse,  $\Delta g$ ($\Delta v \sim  \SI{0.8}{\pico \meter/ \second}$, in Table~\ref{tb:zerocrossing}). This rules out that the spike is in $g_1$, and leaves only the options that it might have been due to a force impulse  on the spacecraft, or to a feature in the $x_{1,\OMS}$ interferometer readout. 

\subsection{Other time series}
We also analyzed the time series of magnetometers  and of the various thermometers of the LTP, without finding any coincidence.

We also inspected the time series of the inertial forces, calculated as explained in \cite{PhysRevLett.120.061101}, and found no coincidence. Note that, as stated in Sec.~\ref{sec:dyn}, these series had been subtracted from the observed acceleration to form $\Delta g$. Thus a true force glitch in any of these series, should have emerged from the original acceleration series,  and should have disappeared in the subtraction. Thus our check was against nonforce data artifacts in the inertial force series that may have been transferred to  $\Delta g$ upon subtraction, or against miscalibration in the subtraction, that would have left a residual glitch in $\Delta g$.

We also inspected the time series of the  forces on the test masses along directions orthogonal to $x$, but the force sensitivity along these axes is too low, with respect to that along $x$, to give any significant information. This reduction stems from both the lack of an interferometric readout on $y$ and $z$, and from these degrees of freedom being used to control the motion of the spacecraft. 

Note that the effects of the acceleration of the spacecraft has already been subtracted from data as illustrated in Sec.~\ref{sec:dyn}. In particular, the  $\delta_{x_1} \ddot{x}_{1,\OMS}$ term in $g_\text{SC}$ in Eq.~\eqref{eq:deltag} subtracts the effect of spacecraft acceleration being picked up by the finite common mode rejection of the differential interferometer, and the term $(\omega_2^2-\omega_1^2) x_{1,\OMS}(t)$ subtracts the effect of  the ``harmonic'' coupling to the spacecraft acceleration.
Thus our inspection of the $x_{1,\OMS}$ series is a search either for possible disturbances affecting simultaneously both $\Delta g$ and $x_{1,\OMS}$, or for incomplete subtraction due to calibration errors.

\section{\label{sec:disc} Discussion}
We discuss here the implications of the observations described so far for the identification of the possible sources of the observed glitches.

Before we start the discussion, it is useful to give some clarifications on the cold runs, during which the instrument was operated well outside its nominal working range. 

Besides the increased glitch rate, we observed two major effects of such nonstandard operation mode. First, all interferometric channels of the OMS were characterized by increased noise and spurious signals. Second, the system was subject to a significant mechanical distortion. 

We could detect such distortion by monitoring the time series $\Delta x_\GRS(t)-\Delta x_\OMS(t)$ that measures,  to first order, the difference between the relative displacement of the two electrode housings to that of the two test masses. 

During cold runs this difference kept changing over time, with a total variation of up to a few \si{\micro\meter}, likely due to the thermal distortion of the mounting structure of the two GRS. This distortion must have put under severe stress the interface between the GRS and the glass structure of the OMS, which, on the contrary, is virtually undistorted by temperature because of its very low temperature coefficient.
\subsection{Fast, low-impulse glitches}
We discuss next the fast, low-impulse glitches. The fast timescale (Fig.~\ref{fig:glitch_eg} right), the absence of any feedback force signal, the absence of any net impulse (Fig.~\ref{fig:deltaThis2sided}, lower panel), and, most importantly, the coincidence with events in other interferometer channels for the majority of them, are all features that  point, for these glitches, to an explanation as interferometer anomalies.

Note that glitches of this kind  were very rare in ordinary conditions (less than one per month), while their production has been boosted by the relatively unstable situation of the  cold runs, in some occasion in the form of clusters that violate the Poisson condition for random occurrence times (Fig.~\ref{fig:deltaThis2sided}, upper panel). 

We were not able to trace back the true  generating mechanism behind these anomalies. It is however worth mentioning that similar events also showed up in a similar interferometer flown on the GRACE Follow-on mission \cite{PhysRevLett.123.031101}. The interferometer readout exhibited phase jumps that would translate for us into steps in $\Delta x$, which were traced back to mechanical disturbances generated by thrusters activation. Thus mechanical stress may be the root cause of these interferometer anomalies on LPF.

Since the mechanism behind these transients is not fully understood, we cannot predict whether they will occur in LISA in a similar fashion. For instance, while LPF interferometry reached a sensitivity $\SI{32}{\femto\meter/\rtHz}$ \cite{PhysRevLett.126.131103}, LISA is expected to be limited at $\SI{10}{\pico\meter/\rtHz}$.

Nevertheless it is worth mentioning that the  science degradation resulting from this kind of glitch, if there were to be any in the LISA data, is not expected to be severe, given their very fast timescale affecting just a few data points, and the lack of a low frequency component due to the corresponding lack of impulse.

In addition we note that the largest of these events in the interferometer outputs might be detectable on ground. Though much noisier than in flight, LPF interferometry could demonstrate on ground $\simeq \si{\pico\meter/\rtHz}$  sensitivities \cite{Ifoground}. This should allow detection and study of, for instance, the largest, tens of \si{pm} high, steps in the interferometer output. The effect of environmental conditions on such glitch production could then be investigated.

\subsection{Impulse-carrying glitches}
Contrary to fast low-impulse glitches, we were able to demonstrate that a large fraction of the impulse-carrying ones consists of true force events acting on one or both test masses. Though we could not extend the demonstration to the smallest ones, the shape similarity with the larger ones, and the continuity of the parameter distribution,  make it very likely that  all glitches of this kind, or at least the vast majority of them, consist of  true force events. 

In the following, after discussing their main statistical features, we discuss the possible sources for such force events, with the aim of ruling out the unlikely ones, and identifying the most likely ones.

\subsubsection{Glitch taxonomy and some implications for their origins}

To gain more insight into the nature of the force glitches, let us now discuss their statistical properties.

First, their  time of occurrence  has been following Poisson statistics  during both ordinary and cold runs (Fig.~\ref{fig:tauhistograms}). We observed  neither any significant clustering nor any repeated pattern. Thus these glitches appear to be due to independent sources and to occur at random times. 

In particular, during ordinary runs in stable conditions and within the specified operating conditions for the LTP,  we observed a constant mean rate of occurrence (Fig.~\ref{fig:ordinaryrate}) throughout the $\simeq \SI{1.2}{\year}$ of the mission science operations. During this same period of time, the pressure around the test mass had been decreasing by  almost an order of magnitude \cite{PhysRevLett.120.061101}, and  many changes of operational settings had been taking place in between any of the different noise runs we have been discussing here, both to maintain the orbit, and to perform dedicated investigations \cite{PhysRevD.97.122002}.

Cooling of the system to near $\SI{0}{\celsius}$ increased this rate by more than one  order of magnitude. However such a rate increase was not uniform across the glitch parameter space.

More specifically, Fig.~\ref{fig:Glitch_LPF_spSNR3} shows two basic features:
\begin{itemize}
    \item a population of glitches with $\Delta \gtrsim \SI{1.6}{ks}$, which accounts for 37\% of  all positive glitches in ordinary runs,  for just 0.7 \%  in cold runs, and is absent, for negative impulse glitches, in all runs;
    \item in both ordinary and cold runs, positive impulse glitches constitute the vast majority of all glitches. For convenience we remind here that a positive impulse pushes the test masses one toward the other.
   \end{itemize}
In Table~\ref{tb:freq} we list the number of glitches in these three categories: ($\Delta v>0$, $\Delta< \SI{1.6}{ks}$), ($\Delta v>0$, $\Delta> \SI{1.6}{ks}$), and ($\Delta v<0$, $\Delta< \SI{1.6}{ks}$), for both ordinary and cold runs. The table also contains the projected counts one would have observed during the \SI{11.9}{\day} long cold runs, had the statistics not been affected by the cooldown (see Appendix~\ref{app:APCR}). 
The result of this  projection shows that:
\begin{itemize}
    \item Counts for positive glitches of duration less than \SI{1.6}{ks} are more than twenty times larger than the largest projected value. Thus cooldown has strongly increased the rate of these glitches.
    \item Counts for positive glitches of duration larger than \SI{1.6}{ks} are compatible with the projected values, and thus the rate of these glitches has not  increased upon cooldown.
    \item Counts for negative glitches, all of which are short,  are more than four times larger than the value expected from the ordinary runs statistics. Thus cooldown has  affected the rate of these glitches too, though to a lesser extent than that of positive shorter glitches.
\end{itemize}

 \renewcommand{\arraystretch}{1.2}
\begin{table}[b]
\caption{\label{tb:freq}
Observed and expected number of glitches listed by duration and impulse sign.}
\begin{ruledtabular}
\begin{tabular}{cccc}
Runs & \makecell{$\Delta v>0$\\$\Delta < \SI{1.6}{ks}$} & \makecell{$\Delta v>0$\\$\Delta > \SI{1.6}{ks}$} & $\Delta v<0$ \\
\hline
ordinary & 51 & 30 & 17\\
cold & 304 & 2 & 28\\
ord.$\to$cold\footnotemark[1] & 1-13 & 0-9 & 0-6\\
\end{tabular}
\end{ruledtabular}
\footnotetext[1]{Counts for cold runs, projected from the observed counts and rate in ordinary runs. Intervals correspond to 90\% confidence.}
\end{table}

We have also compared  the  ordinary runs glitch parameter distributions with that for cold runs, for the categories of glitches that are  significantly populated in both type of runs. 

Statistical tests on the equality of two multivariate distributions are still subject to a debate that goes well beyond the scope of this paper. To get nevertheless a sense of the similarity between the  two distributions, we have performed marginal Kolmogorov-Smirnov tests for $\Delta$, $\Delta v$, and  $\dgmax$. This last parameter is $\propto \left|\Delta v\right|/\Delta$, and thus mixes somewhat the populations of the two other parameters. The resulting $p$-values are reported in Table~\ref{tb:KS}.
 \renewcommand{\arraystretch}{1.2}
\begin{table}[b]
\caption{\label{tb:KS}
Resulting $p$-values from the marginal Kolmogorov-Smirnov test, for parameter distributions in ordinary and cold runs.}
\begin{ruledtabular}
\begin{tabular}{cccc}
Category & $p_{\Delta}$ & $p_{\Delta v}$ & $p_{\dgmax}$ \\
\hline
$\Delta v>0,\,\Delta \le \SI{1.6}{ks}$ & 0.26 & 0.33 & 0.31\\
$\Delta v<0$ & 0.35 & 0.28 & 0.28\\
\end{tabular}
\end{ruledtabular}
\end{table}

Though these considerations are not at all a proof, they are nevertheless suggestive that in ordinary runs, the glitches belong to three different families. 

\begin{itemize}
    \item The first is a family of positive impulse glitches with $\Delta \ge \SI{1.6}{ks}$. 
    
    For this family $\Delta v$ increases rapidly with $\Delta$. For reference, we calculate that the  line $\Delta v \, [\si{pm/s}]=\Delta^2 / \left(\SI{0.65}{ks}\right)^2$ is the lowest power-law upper bound to all  glitches in this family.
    
    Though the shape of distribution might be affected by  the detection threshold from below, such  quadratic upper bound does not appear to be  due to any obvious  selection effect.
    
    Remarkably, the rate of these glitches has not been affected by the thermomechanical stress conditions of the cooldown. 
    
    As the threshold $\Delta \ge \SI{1.6}{ks}$ has no particular physical meaning, this family may also include some of the shorter duration glitches. Actually if the quadratic upper bound above  were used to define the family, the family would  include 3 more glitches in ordinary runs and 2 more in cold runs. However the conclusions about the count projections would not be modified by these adjustments.

    \item The second family is composed of positive $\Delta v$ glitches with $\Delta < \SI{1.6}{ks}$. This coincides basically with the   glitch population of this kind during the cold runs.
    
    Indeed, given that the rate of the glitches of the long positive impulse family was unaffected by cooldown, even if that family extended to shorter duration, very few samples would contaminate the cold runs distribution with $\Delta < \SI{1.6}{ks}$.

    About 70\% of the glitches within this family have duration less than one minute. However the family also includes a sparse sample of glitches with duration that can take values up to just below the $\Delta = \SI{1.6}{ks}$ threshold.
    
    The rate of these glitches has been greatly affected by the thermomechanical conditions of cooldown. Actually, as in cold runs  negative impulse glitches are only about 8\% of total, the  time/temperature evolution of the rate in Fig.~\ref{fig:coldseries} refers basically to the glitches in this family.

    \item Finally, the third is the relatively small family  of negative impulse glitches, again  with $\Delta \le \SI{1.6}{ks}$. 
    
    About 53\% of these have duration of less than a minute, and these too include samples with longer duration, approaching $\Delta = \SI{1.6}{ks}$. 
    
    The rate of this family  has also been affected by cooldown, though to a lesser extent, but there are not enough samples to assess if the rate has been affected by temperature. 
    
\end{itemize}

\subsubsection{Possible physical sources of force glitches}

In the light of all the evidence above, we now discuss the possible physical sources of these force glitches.

\vspace{\baselineskip}
\noindent\textit{Platform acceleration and inertial forces.}---An obvious cause of transient ``events'', in space-borne differential accelerometers, may be some corresponding events in the acceleration of the spacecraft, which would be picked up because of  the finite common mode rejection of the instrument. An obvious example of this is the already mentioned case of a micrometeoroid hit.

In our case we can rule out this source, as we have  been correcting the data, as described in Sec.~\ref{sec:dyn}, for the coupling to  spacecraft motion. In addition we  have inspected the $x_{1,\OMS}(t)$ data series, to check for any residual coincidence, possibly due to some residual inaccuracy in the data correction, finding none. Similar data correction and inspection  also rule out, as sources of glitches, inertial forces due to spacecraft rotation.  

\vspace{\baselineskip}
\noindent\textit{Thermal effects.}---Also the lack of coincidence with thermometer readings allows us to rule out some possible explanations. 

The correlation of temperature and temperature gradient variations with $\Delta g$ has been investigated by a series of dedicated experiments during the operation of LPF \cite{thermal-paper-rita}. We have measured the dependence of $\Delta g$ on the average temperature $\overline{T}$ of all thermometers on each electrode housing. Upon heating the electrode housing we found a complex behavior with a relatively prompt response and coefficient $\partial \Delta g/\partial \overline{T}\vert_{p}$, plus a delayed response with  coefficient $\partial \Delta g/\partial \overline{T}\vert_{d}$, likely due to the delayed heating of distant sources. The former was found to be pretty constant over time, at about $\SI{0.5}{\pico\meter\,\second^{-2}\,\kelvin^{-1}}$. The coefficient $\partial \Delta g/\partial \overline{T}\vert_{d}$ was instead found to decrease by about a factor 5 over the course of the mission and to level off at $\simeq \SI{0.5}{\pico\meter\,\second^{-2}\,\kelvin^{-1}}$, paralleling the decrease in pressure. 

A similar pressure-dependent behavior was also, as expected,  found for $\partial \Delta g/\partial \Delta T_{\text{EH}_i}$, where $\Delta T_{\text{EH}_i}$ is the difference of temperature across the $i$th electrode housing. For both electrode housings, these coefficients leveled off at $\simeq \SI{10}{\pico\meter\,\second^{-2}\,\kelvin^{-1}}$. 

With such sensitivities, to explain the smallest of the observed $\Delta g$ glitches with a  glitch in $\overline{T}$, one would need amplitudes of order \si{\milli\kelvin} at the beginning of mission and of many tens of them at the end. We would have detected such glitches, as our resolution is of order of tens \si{\micro\kelvin} in $\overline{T}$ for a \SI{100}{s} glitch following the template in Eq.~\eqref{eq:onesidedshape}.

The same applies to a glitch  in the differential temperature $\Delta T_{\text{EH}}$, for which glitch amplitudes would need again to be of order of \si{\milli\kelvin}, and where our sensitivity is in the \si{\micro\kelvin} range thanks to a dedicated low noise temperature differential readout  \cite{10.1093/mnras/stz1017}.

Therefore such hypothetical temperature glitches would have been detected within their relative time series. We believe that this rules out the hypothesis that glitches may be due to thermal transients in the system.

\vspace{\baselineskip}
\noindent\textit{Gravitational signals.}---The lack of any significant permanent change in the force, upon the occurrence of any of these force events, rules out the possibility that they may consist of the gravitational signals from some amount of mass permanently changing position or leaving the system. This would be the case, for instance,  for a large outgassing event from some point of the spacecraft or for some sudden leak of a large amount of propellant.

\begin{figure}[htbp!]
  \centering
  \includegraphics[width=\columnwidth]{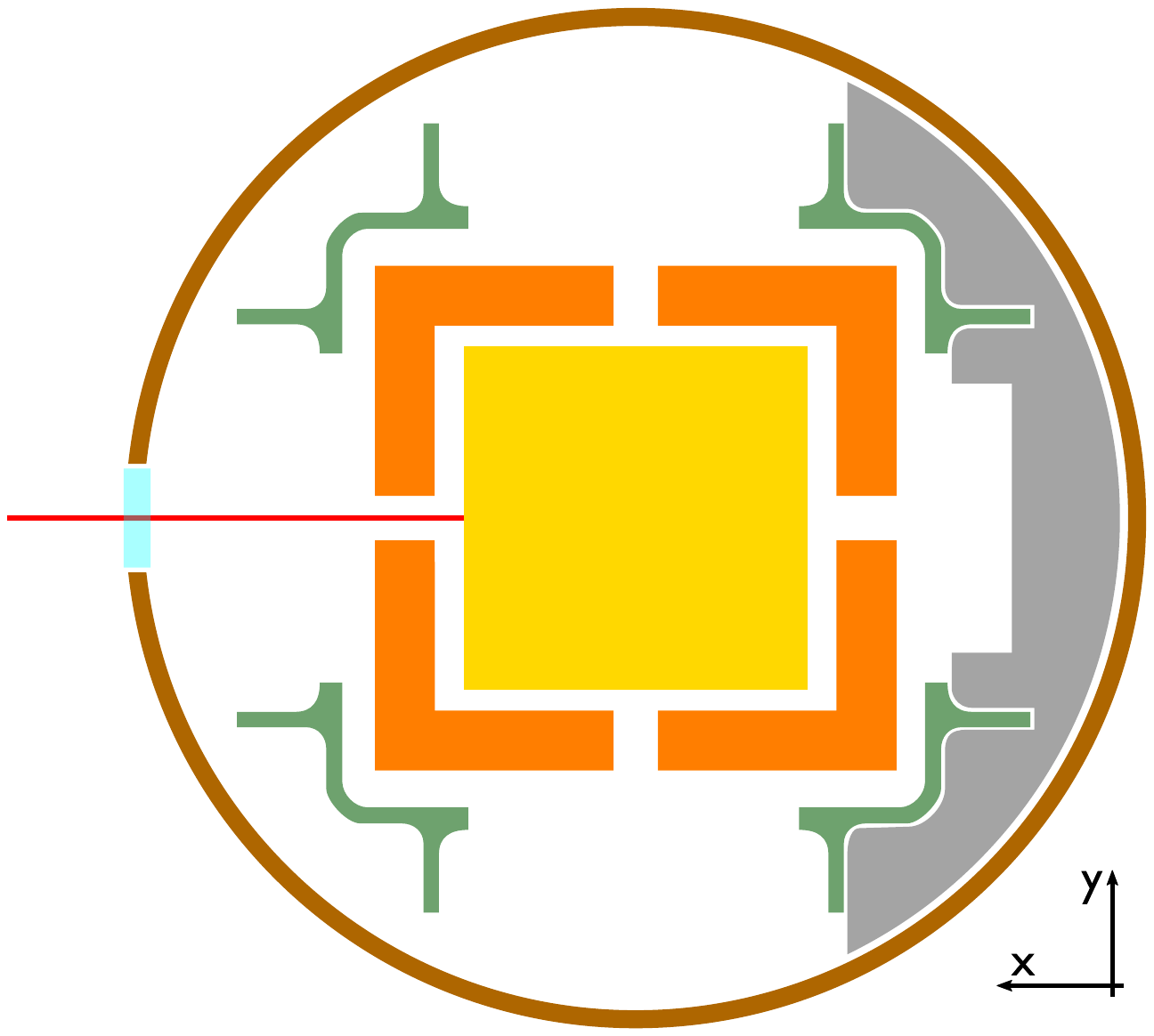}
  \caption{Schematic representation of the  GRS, horizontal section. The yellow square is the  Au-Pt test mass. The orange hollow square represents the electrode housing, whose mid section carries four symmetric holes, one of which is the input port for the laser beam (in red). The green part is the section of a titanium  structure that holds the various elements together, while in gray is the specially shaped tungsten gravitational  balance mass. The brown circle is the section of the vacuum chamber, while the cyan rectangle represent the optical window to transmit the laser beam.} 
  \label{fig:TM_EH_YS_IBM}
\end{figure}

A different origin of a gravitational signal would be a body reversibly moving around its equilibrium position. An example of that is the \SI{2}{\kilogram} tungsten mass located within each GRS, at a few centimeters from the test mass center to suppress the static gravitational field on the test mass \cite{gravity} (see Fig.~\ref{fig:TM_EH_YS_IBM}).

The gravitational force gradient per unit mass acting on the test mass due the balance mass is $-\omega_{1,2}^2\simeq \SI{5e-7}{s^{-2}}$ (see Sec.~\ref{sec:dyn}) \cite{gravOHB}. Thus, to produce our smallest glitches with a peak amplitude about $\si{fm\,s^{-2}}$, one would need a displacement of the balance mass out of its equilibrium position peaking at $\sim \SI{2}{nm}$ and then getting back to rest.

The balance mass is the largest source of gravitational gradient, the gradient from other sources being significantly smaller, as the gradient from the rest of the GRS drops to about $\SI{4e-7}{s^{-2}}$, due to compensation \cite{gravOHB}, and as the contribution of farther apart components decays as the cube of the distance. For instance, the optical bench contributes a gradient of about $ \SI{6e-8}{s^{-2}}$, the other GRS  $\SI{3e-8}{s^{-2}}$, and  the rest of spacecraft  $\SI{1e-8}{s^{-2}}$ \cite{gravASU}.

Characteristic frequencies of mechanical parts surrounding the test mass are in the kHz range. Thus, even the over damped mechanical motion of those parts takes place on timescales much shorter than that of the vast majority of the observed glitches. In order to produce a glitch by way of this physical mechanism, one would need parts to move because of some transient thermomechanical distortion, and not by some mechanically excited free motion.

We foresee two major damaging patterns for such a thermomechanical distortion. The first is an expansion or contraction of the GRS around the test mass due to a temperature fluctuation. The major effect of such distortion would be to move the balance mass relative to the test mass. Given the construction details, this effect adds a term of order $\simeq \SI{0.3}{\pico\meter\,\second^{-2}\,\kelvin^{-1}}$ to the temperature coefficient $\partial \Delta g/\partial \overline{T}\vert_{p}$ discussed above. The possibility of glitches originating this way has been already ruled out.

The second distortion pattern that may give origin to a force transient, is a rigid displacement along $x$ of either of the two GRS relative to its own test mass. This is the case, for instance, if the struts that attach the GRS to the LTP bay thermally expand or contract, moving both GRS in opposite directions, and changing their distance by some amount $\delta l$. 

By moving the main sources of gradient, such distortion would cause both a signal $\Delta g=-\omega_{2,\GRS}^2 \, \delta l$, with $-\omega_{2,\GRS}^2\sim -\omega_{2}^2 \sim \SI{4e-7}{s^{-2}}$, and  a signal $x_{\OMS}-x_{\GRS}=\delta l$. As already mentioned, we have observed both such signals  upon the large distortion caused by  cooldown.

Thus, a glitch originating from such a distortion pattern, or from any pattern that would move any of the GRS relative to its test mass, should also show up in the $x_{\OMS}-x_{\GRS}$ series, peaking at $\dgmax/\left(-\omega_{2,\GRS}^2\right)$, a number that ranges from a few nanometers to micrometers.  As discussed in Sec.~\ref{sec:grsoms}, we have inspected the series $x_{\OMS}-x_{\GRS}$ and found no corresponding glitch. 

A distortion pattern moving the spacecraft, or one of its large components, relative to the entire LTP and nondetectable  in $x_{\OMS}-x_{\GRS}$ would require much larger amplitudes. The difference of gradient on the two test masses due to the entire spacecraft, which is the quantity  that would matter in this case, is of order $ 10^{-9}~\si{s^{-2}}$. Thus, to produce the observed glitches, the spacecraft would have to have moved along $x$ by an amount ranging from micrometers to millimeters, requiring temperature changes ranging from a fraction of a \si{\kelvin} to hundreds of \si{\kelvin}, without causing any detectable distortion within the LTP. We consider such a scenario to be quite unlikely. 

In conclusion, we believe that a gravitational origin is an unlikely explanation for the vast majority of the observed glitches.

\vspace{\baselineskip}
\noindent\textit{Magnetic force.}---It also appears unlikely that these force glitches are explained by some slow transient in the magnetic field.  With \emph{slow} here we mean that we are not considering eddy current effects, which we will  discuss later.

The magnetic susceptibility of the test masses has been measured to be $\chi\sim\num{3e-5}$ \cite{labtoLPF} and its permanent magnetic moment $|\bm{\mu}|< \SI{5}{\nano\ampere\meter^2}$ \cite{magOHB}, though this is just an upper limit. The static magnetic field on board LPF was found to be $|\bm{B}|\sim \SI{1}{\micro\tesla}$ \cite{magnetic-mnras}, and, finally, from the lack of correlation between the magnetic field and force noise \cite{lpf_noiseperf3}, we estimate the magnetic gradient to be less than $\SI{10}{\micro\tesla/\meter}$. 

The force peak amplitudes of all observed glitches have $\dgmax \ge 10^{-15}~\si{\meter\,\second^{-2}}$ (see Fig.~\ref{fig:peakhis}). 
To reach such a force level, a glitch in the magnetic field gradient, on either of the test masses, considering only the coupling to the test mass induced moment, should peak at about $\SI{1}{\micro\tesla/\meter}$. Unless we take into account an unrealistically close  source, a gradient like this would have produced a detectable signal on some of the magnetometers, at least for glitch duration larger than about \SI{10}{s}. As glitch amplitudes were even 1000 times larger than this minimal one, many of the glitches would have produced large magnetometer signals. 

Similar conclusions are obtained assuming that the glitch is in the magnetic field.  To produce our force glitch with a glitch in the magnetic field, one would require a peak at least $\SI{0.1}{\micro\tesla}$, well above the detection threshold. Thus the lack of observed magnetic glitches rules out this explanation.

A further mechanism for force transient of magnetic origin, is that of currents induced  via Seebeck effect by thermal gradients inside the test mass. The test mass material has a finite amount of Au and Pt precipitate that may create effective thermocouples  at grain boundaries. The effect was noticed during magnetic characterization of the test masses on ground, during which temperature differences of order of a few  \si{\kelvin} across the test mass, due to manipulation by human hands, induced a magnetic moment peaking at some \si{\micro\ampere\square\meter}.  With the value for the static gradient quoted above, this effect would just give a correction of  $\simeq \SI{0.1}{\pico \meter\,\second^{-2}\,\kelvin^{-1}}$ to the coefficient $\partial \Delta g/\partial \Delta T_{\text{EH}_i}$ mentioned in the previous section, and its role in producing glitches is ruled out together with the rest of the thermal effects.

In addition to these low frequency effects,  magnetic fields at high frequency may induce eddy currents within the test masses and then exert Lorentz forces on them \cite{labtoLPF}. The effect is thus quadratic and would convert the low frequency amplitude fluctuations of a high frequency magnetic spectral line into a corresponding low frequency force. 

To give a scale of the effect,  a recent finite-element  electromagnetic  calculation by the LISA project \cite{eddy}, has shown that the effect of a dipole of $\SI{1}{\milli\ampere\,\meter^2}$ located at a distance $d=\SI{20}{cm}$ from the test mass and oscillating at the frequency of 100 Hz, would cause a force of $\Delta g \simeq \SI{4}{\femto\meter/\second^2}$. The effect reaches its peak at \SI{100}{\hertz}, while at lower frequency the induced current decreases and above that the screening effect of the metallic electrode housing attenuates the oscillating field. 

The effect of a dipole source decreases with $d^{-7}$, so that at the closest distances of about $\SI{0.4}{m}$ between the test mass and any active device on the LPF spacecraft the effect might be $\sim 100$ times smaller.

The spacecraft prime contractor performed a test campaign on ground  against audio frequency magnetic lines \cite{acmagnetics} during LPF development. A few lines have  been identified with peak amplitudes $< \SI{1}{\nano\tesla}$ at the position of the test masses. In the point dipole model at a distance of $d\simeq \SI{0.4}{\meter}$, each line would be generated by a dipole of $\simeq\SI{0.3}{\milli\ampere\,\meter^2}$ and would exert a static force $\Delta g \simeq \SI{4e-3}{fm\,s^{-2}}$. Even a glitch consisting of 100\% amplitude modulation of any these lines -- a behavior not observed during test --  would have then an amplitude orders of magnitude less than those in Fig.~\ref{fig:peakhis}. 
However, we had no magnetometer on board sensitive to the audio band and thus we cannot exclude that additional, more intense, amplitude modulated lines had been generated once on orbit, as  the operating conditions  may have been significantly different from those during testing.

Also, the shape and timescale of the glitches are not easily reconciled with such an ac magnetic origin. Electromagnetic emission from electronics is usually modulated by noise, by switching among different operational settings, and by thermal variations, and we do not see how these may easily follow  the reversible exponential behavior lasting minutes to hours that would be required to generate the observed glitches. 

One way though of producing a smooth time evolution  is that of two lines of constant amplitude, the frequencies of which would slowly drift over time. If during some time interval these lines had a substantial overlap in the frequency domain, they would indeed generate a force on the test mass.

Lines observed during testing where stable in frequency, but, again, we cannot exclude that other lines were present in flight. It seems, however, highly unrealistic that on orbit enough lines have been generated, with different enough drift rates and shapes, to explain the hundreds of glitches of Fig.~\ref{fig:Glitch_LPF_spSNR3}, with parameter values that span a few orders of magnitudes. 

In conclusion, we believe that while the possibility that some of the observed glitches are due to eddy currents cannot be discarded, it is highly unlikely that this source may explain the majority of the observed glitches. Nevertheless we certainly recommend that in LISA a thorough testing is performed on ground, and that on-board diagnostic magnetometers with sensitivity up to \SI{1}{\kilo\hertz} are considered. 

\vspace{\baselineskip}
\noindent\textit{Electrostatic forces from GRS electronics}---Each LPF test mass is electrically isolated from its surrounding with no detectable discharging path. Thus an  event of charging because of cosmic rays \cite{PhysRevLett.118.171101,armano_characteristics_2018} or any other source of particles, would show up as a step in $\Delta g$, quite incompatible with the observed finite-impulse glitches.

In addition, all surfaces facing the test masses are conducting and grounded, and would not accumulate free charge.

Still, glitches may be produced by spurious voltage transients in the electronics we use to control the test masses. More specifically, the mentioned electrodes facing the $x$-faces of the test masses, are all driven by separate amplifiers. A voltage  glitch in one of these amplifiers would certainly produce a force on the test mass.

However, a single-electrode event like this would also generate a torque around $z$, with a lever arm $|r_\phi|=\SI{11}{mm}$, which is the main reason why we have searched for a torque component to the detected force glitches as explained in  Sec.~\ref{sec:torque}. 

To span the observed range of glitch peak amplitudes, from \SI{1}{fm/s^2} to \SI{1}{pm/s^2}, with the electrode geometry, there would need to be a transient change in the mean square voltage at the actuation amplifier output between roughly \SI{10}{mV^2} and \SI{10000}{mV^2}. This could occur in different forms:
\begin{itemize}
\item A transient ``quasi-DC'' voltage in the \SI{100}{\micro\volt} to \SI{100}{\milli\volt} range, mixing with stray DC potential differences of order \SI{100}{\milli\volt}, due to test mass charge and/or stray ``patch'' voltages. 
\item A transient change in the roughly \SI{1}{V} actuation audio-frequency carriers \cite{armano:subfemtog} by roughly \SI{10}{\micro\volt} to \SI{10}{\milli\volt}.
\item A transient electrode oscillation coherent with the \SI{100}{\kilo\hertz} sensing ``injection'' frequency \cite{PRD_96_062004_capacitive} and mixing with the \SI{0.6}{V} amplitude test mass bias, in roughly this same amplitude range. 
\item A spontaneous AC oscillation, at some random frequency not associated with the actuation or injection, with amplitude in the \SIrange{5}{200}{\milli\volt} range.
\end{itemize}
While we cannot directly exclude any of these -- though the \SI{100}{\kilo\hertz} excitation would have likely given some capacitive sensing error -- they were not detected in dedicated preflight tests, albeit relatively short (less than day per electrode), which could have detected such anomalies.  Certainly such features can and should be investigated more thoroughly on ground for the  LISA electronics.

A strong indicator that the glitches do not originate in the actuation electronics comes from the analysis of the possible torque component to the observed glitches.
The findings in Sec.~\ref{sec:torque} show that:
\begin{itemize}
     \item There are in total 56 glitches within ordinary runs, spread  over the entire parameter space,  for which we would have been able to detect a  lever arm  of $\SI{11}{mm}$ (see Fig.~\ref{fig:arm}). Only one of these is both incompatible with  $r_\phi= 0$ and compatible with $r_\phi=\SI{11}{mm}$ (see Table~\ref{tb:eventtorque}). Though there is no proof that this glitch is indeed of electrical origin, one might nevertheless take 1/56 as a rough bound to the fraction of glitches that may be due to this source. 
    \item For the cold runs glitch data, none of the 147 glitches for which there is sufficient resolution to resolve $r_\phi=\SI{11}{mm}$ have such an effective arm. The probability of such an event, using binomial statistics, and assuming the distribution is the same as during ordinary runs, is $p=0.08$, a figure that does not allow us to reject the equal distribution hypothesis. Using both observations, ordinary and cold runs, the probability of such an occurrence becomes $p\le0.023$ with 95\% confidence.
    \item For the kind of standard, audio-frequency electronics we are discussing here, minute to hour long transients which are not induced by some corresponding thermal transients, are quite unexpected. Of the 121 glitches with $\Delta \le \SI{1}{\minute}$ and detectable $\SI{11}{mm}$ lever arm, none is found to have such a lever arm, which  gives $p\le0.024$ with 95\% confidence. 
\end{itemize}

While this effective arm test is inconclusive for the smallest and fastest glitches, for which our sensitivity to a lever arm is reduced, most of our glitches are incompatible with a single-electrode electrical origin.

In addition to this, even for the smaller, faster glitches, the observed increase in rate upon cooling the spacecraft is not easily reconciled with an electrical origin.

More complex voltage events, simultaneously affecting more than one electrode -- such as two adjacent electrodes which combine to give force without torque -- are even less likely given the design of the electronics \cite{Neda2020}. 

Testing for the different types of transient voltages that could produce glitches at the levels and rates observed in LPF would require dedicated detection circuitry and long measurements.  The voltage levels are however accessible, and such testing in preparation for LISA is recommended.

\vspace{\baselineskip}
\noindent\textit{Outgassing environment.}---One candidate source of force is the exchange of momentum between the test mass and the gas molecules surrounding it. This exchange, in the form of Brownian noise, dominated the noise budget at frequencies above about \SI{1}{\milli\hertz} \cite{armano:subfemtog}. 
 
 Gas pressure around the test mass, that we deduced from the Brownian noise, decayed over the course of the mission, as the vacuum chambers were vented to space via venting ducts. Pressure went from  about $\SI{10}{\micro\pascal}$ at the beginning of the mission, to about $\SI{1}{\micro\pascal}$ toward the end, following a power law function of time, strongly indicative of water outgassing \cite{lpf_noiseperf3}.
 
 The vacuum environment of the test mass is rather unusual, as the vacuum chamber is densely packed with components: test mass, electrode housing, test mass launch-lock mechanism, various cable bundles, etc. Thus, the outgassing surface to volume ratio is unusually high for a vacuum system, and the distribution of outgassing surfaces rather non symmetric around the test mass.

In such an environment, one possible source of glitches may be an event of release of some metastably trapped gas from pores. Similar events are often observed in vacuum systems, due to so-called virtual leaks -- cavities  with a high impedance connection to the outside -- that may trap gas  and release it in bursts \cite{hoffman_handbook_1998}. The phenomenon is also known to be triggered by mechanical stress and friction.

\begin{figure}[htbp!]
    \centering
    \includegraphics[width=\columnwidth]{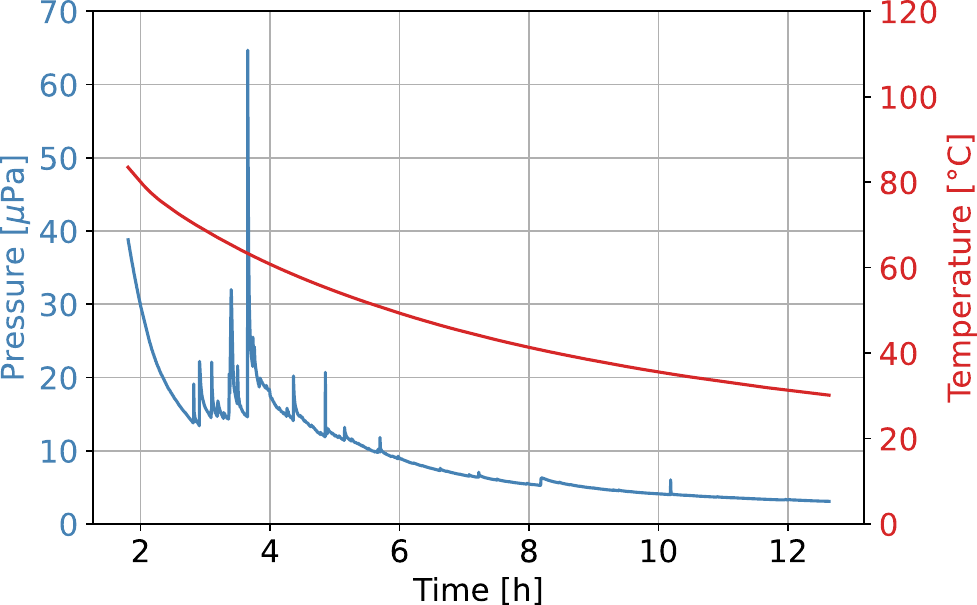}
    \caption{Blue line and left scale: pressure in one of the two  GRS during cooldown, after about $\SI{24}{\hour}$ pumping at about $\SI{115}{\celsius}$ for vacuum preparation (bakeout) on ground. Red line and right scale: temperature in the test facility. Vacuum preparations were performed by inserting the GRS, with its venting valve open, inside a wider vacuum chamber. The pressure and temperature shown in the figure are those of this wider chamber. The time origin is set at the end of the bakeout phase.  (Data courtesy OHB Italia)}
    \label{fig:bakeout}
\end{figure}

We have observed events similar to glitches in the pressure gauge time series during  vacuum preparation of the GRS on the ground (see Fig.~\ref{fig:bakeout}). For that test the GRS had been inserted in a wider vacuum chamber with its venting valve open. The chamber was pumped down and its temperature  was raised to and maintained at $\simeq \SI{115}{\celsius}$ for about \SI{24}{\hour} to get rid of most of the adsorbed water, a standard procedure for vacuum systems known as bakeout.

The stretch containing the glitches in Fig.~\ref{fig:bakeout} was observed during final cooldown, during which the system was subject to a significant amount of thermomechanical stress due to the relatively rapid contraction. A similar behavior with many spikes was also observed during the preparation of the other GRS, while no spikes were detected during a test with the empty chamber. However, we have no way of assigning with certainty the source of these gas emission spikes to the GRS interior, and we only show them here as an example of the phenomenon in vacuum systems.

Molecular simulations show that a molecule released from a surface nearby the test mass can transfer a net momentum to it before leaving the system via the venting duct. For instance, using the same simulation method and  simplified geometry described in \cite{PhysRevLett.103.140601,thermal2007}, we calculate that a water molecule with a Maxwell-distributed momentum, entering the laser port in the center of the $x$-face of the electrode housing (Fig.~\ref{fig:LPFfigure}, \ref{fig:TM_EH_YS_IBM}) and hitting the test mass,  would exchange with the test mass an average momentum per unit mass along $x$ of $\Delta v\simeq \SI{-2.4e-22}{m/s}$, before leaving the electrode housing through any of its holes. This is about a factor 40 larger than the mean momentum of the distribution  $\left< \Delta v\right>\simeq \SI{5.7e-24}{m/s}$, an enhancement due  to  multiple collisions between molecule and  test mass, caused by the constrained geometry \cite{PhysRevLett.103.140601}. 

Using the figure above for the impulse per molecule: the glitch with largest test mass impulse $\Delta v \simeq \SI{1}{nm/s}$ would correspond to roughly \num{4e12} water molecules ($\SI{0.13}{\nano\gram}$) hitting the test mass; the sum of all glitches in ordinary runs and  during the entire mission, assuming the observed constant rate,  would amount to $\simeq \SI{1.5}{\nano\gram}$; those observed during cold runs would amount to $\simeq \SI{0.15}{\nano\gram}$. These figures constitute a comparatively small amount of molecules,  when  compared with the  total outgassing rate from each GRS, of the order of $\SI{100}{\micro\gram/\day}$ at the beginning of the mission and $\SI{10}{\micro\gram/\day}$ at the end.

We have also performed some molecular dynamics simulations with \textsc{Molflow} \cite{Molflow}, on a more realistic model of the GRS. This model includes the real geometry of the electrode housing, the tungsten balance mass, the cables, and many other details. The tool cannot calculate the total momentum transferred to the test mass, however it calculates the differential pressure between its opposite $x$-faces. This is a reasonable proxy for the momentum transferred per unit time, though  certainly underestimated, as it neglects the momentum along $x$ transferred to the $y$ and $z$ faces. 

We have simulated, as an example,  an instantaneous emission of molecules from a point  on the side of the tungsten mass that faces the aperture in the center of the $x$-face of the electrode housing, identical and opposite to the mentioned laser port (see Fig.~\ref{fig:TM_EH_YS_IBM}).  This emission indeed creates a glitch in the differential pressure with shape similar to the observed ones. In particular the profile never crosses the line $\Delta g=0$. The timescale and the details of the profile however depend on the assumed sojourn time of the molecules on the various surfaces, and on other assumptions in the model. 

What is independent of these details is the total transferred momentum $\Delta v$, that we calculate by integrating the force time profile. With $m_\text{m}$ the total mass of the emitted molecules, we find  $\Delta v /m_\text{m} \sim \SI{0.5}{\nano\meter\, \second^{-1}\,\nano\gram^{-1}}$. Note that the simulation shows that only 20\% of the molecules emitted from the balance mass indeed enters the electrode housing, while the others follow different paths. Thus, given also the intrinsic momentum underestimate of \textsc{Molflow}, this result is not inconsistent with that of the  simplified simulation above.

The tungsten balance masses are a natural candidate for such gas-burst events. First, a microscope analysis has shown that its sintered material is porous, with micrometer size pores \cite{EleonoraPhD}. Second, the sign of the transferred impulse for this source would be positive, as for the great majority of the observed glitches.

Many other components may also trap gas, beginning with the various  bundles of cables that connect the electrodes, and the various motors of the launch lock, to their respective electronics. Some of these sources have the proper position to create also negative impulse glitches that indeed we are able to reproduce with  \textsc{Molflow}. 

Note also that simulations show that the lever arm of a gas inflow from the main inlets to the electrode housing, like that coming from the balance mass, is negligible. Thus, for  the few observed glitches with nonzero lever arm, one should assume that the gas has been emitted by some source localized inside the electrode housing.

However, no other source creates the same sort of cavity around one of the electrode housing apertures as that created by the special shape of the tungsten balance mass. This is reflected in the fact that,  for all other sources, the ratio between the number of molecules hitting the test mass and that of those following different paths is always significantly smaller than that for the tungsten balance mass.

One more argument that may support the gas-release interpretation of glitches is the sensitivity of gas emission to thermomechanical stress, as is well illustrated by Fig.~\ref{fig:bakeout}. As said, thermomechanical stress accompanied the cold runs and their highly increased glitch rate.

In particular, Fig.~\ref{fig:coldseries} shows that the occurrence rate switched almost reversibly, from the $\lambda\sim\SI{1}{d^{-1}}$ of ordinary runs, to the many tens per day of the cold runs, when crossing a comparatively narrow   temperature  range of a few degrees. In addition the data are also suggestive of some slow transient relaxation at the lowest temperatures.  This may indicate that the rate may be following some complex,  non linear stress pattern  due to the significant differential thermal contraction of the high numbers  of equally complex contact interfaces within the GRS.

Though  the likelihood of this source looks the highest among those discussed so far, a few  aspects remain to  be clarified.
An important one  is  the time profile of the glitches and the associated  wide distribution of $\Delta$. Most of the spikes in the data of  Fig.~\ref{fig:bakeout} have an almost instantaneous onset, followed by the  decay pattern one would expect in a standard vacuum system:  large gas releases saturate the adsorption speed of the  chamber solid surfaces and are quickly pumped down, while smaller ones follow a slower decay in quasiequilibrium with surface adsorption. This is the response one would expect for a fast, virtually instantaneous, release of gas from some pocket.

Our simulations show that  the time profile of $\Delta g (t)$, in the case  of  an instantaneous  emission of a group of molecules from a specific source,  also consists of a rise followed by a decay. The rise time is due to the distribution of diffusion times from the source to the electrode housing inlet.  The decay is due to the diffusion of molecules, in the space between electrode housing and test mass, from the inlet to the final exit aperture. The time constant of such decay is substantially fixed and  independent of the rise time, even for molecules  emitted from inside the electrode housing. 

The timescales of both these branches  depend on the assumed sojourn time of the molecules on the various surfaces they encounter along their path. For instance the decaying branch is well fitted by an exponential with a time constant $\sim 45$ times the mean sojourn time of the inside of the electrode housing.

Sojourn times of molecules on metal surfaces depend exponentially on their binding energy. Thus they may vary by orders of magnitude, from $10^{-12}~\si{s}$ to more than seconds, depending on the nature and state of the surface, the nature and amount of adsorbed species etc. It is not easy then to find reliable estimates for a specific situation. As the range of possible values is rather wide, by properly selecting the sojourn times in the model, we have been able to reasonably match, in our simulations, the observed shapes of Eq.~\eqref{eq:onesidedshape_2tau} or~\eqref{eq:onesidedshape}.

However, once the sojourn time for the electrode housing has been selected, the resulting duration for such particular match is fixed. Indeed,  the duration of the decay branch depends only on that choice, and for  $\sim90\%$ of the observed glitches the template is that in Eq.~\eqref{eq:onesidedshape}, which only contains one time constant, so that the duration of the decay branch fixes the overall duration. The model of an instantaneous release of gas from one source  would then not reproduce the observed large variability of glitch duration.

The consequence of this is that, if glitches are due to gas release,  for a large fraction of them their time profile must be dictated by the intrinsic time evolution of the gas release, while it can be limited by diffusion across the GRS only for some of the shortest ones.

We note that standard Fickian diffusion in simple geometries, like from the bottom of a very narrow pit or from the center of a spherical piece of material, does produce a time evolution of the gas outflow very close to that of our observed glitches, and that the  timescale for diffusion may  indeed be very long, depending on the gas species and the material.

However, we were not able to find in the literature any reference to events of slow  gas release from pores or other imperfections. Though there are many qualitative reference to condensation of gas in such  kind of defects, to the possibility that it gets released during pump-down, and to slow, diffusion limited gas motion in porous media, we were not able to find a specific measurement on single events showing such slow  time evolution. The only hint we have of the possibility of some non instantaneous gas evolution is in very few  of the peaks in   Fig.~\ref{fig:bakeout}, showing  indeed some minute-long rise times.

It must be noted though that the scale of the events in Fig.~\ref{fig:bakeout} is orders of magnitude different in amplitude from the kind of release that would explain our glitches. For instance, the above mentioned \texttt{Molflow} simulation shows  that the $\simeq \SI{2}{\nano\gram}$ of total water molecules emitted from the balance mass, needed to generate the  glitch with the largest $\Delta v $, would generate a peak pressure of a few \si{\nano\pascal} at the venting valve, with some reasonable assumption for the sojourn times of molecules, well below the measurement resolution of the figure. 

Additionally, we were not able to find any sound explanation for  the near-quadratic dependence of the impulse, and hence the number of molecules that have hit the test mass, on the duration of the event. 
Fickian diffusion timescales with the square of the length of the diffusion path, with no explicit dependence on the size of the fluid volume that is diffusing.

As a final difficulty for this interpretation, the 3 glitches in Table~\ref{tb:zerocrossing} that cross zero, and have non negligible impulse, are hard to fit to this picture, as none of our molecular simulations could reproduce a glitch for which $\Delta g$ crosses zero. 

The two with smallest impulse  may still be compatible with some interferometer artifact, as we don't have the resolution in $\Delta x_\GRS$ to discriminate against such a case. 

The smallest one might also be of electrical origin, as we do not have the resolution to evaluate the proper value of the lever arm.

The largest impulse one, however, is incompatible with both those options, and would hence remain unexplained by any of the mechanisms we have considered in this section. The observation of a simultaneous event in the $x_1$ interferometer, never observed for all other  impulse carrying glitches, may indeed indicate a different phenomenon, for which we do not have any reasonable model so far. It is worth reminding that leakage of the spacecraft acceleration into $\Delta g$ has been corrected for, and, in this specific case, the correction was anyway negligible.

The possibility that the glitches have their origin in outgassing, exacerbated by mechanical stress, suggests careful avoidance of thermomechanical stress, including operation of the instrument near its integration temperature.  Additionally, while the possible outgassing origin requires specific increased attention to the relevant procedures and testing of LISA GRS hardware, it also suggests caution, with additional testing and analysis, in considering any possible design changes.   

Even more desirable is a dedicated experimental campaign to study if these gas release events exist for the kind of surfaces and elements that compose the GRS interior. Though the amount of gas released is of order of nanograms, its detection as a pressure transient in a properly designed  vacuum system, with a high sensitivity pressure gauge or mass spectrometer, does not seem out of reach. In addition a dedicated experiment is also possible  with the torsion pendulums used to test small forces on the test mass of LPF and LISA  \cite{PhysRevLett.103.140601}. Such experiments are currently under study.

\section{Conclusions} \label{sec:conclusion}
In summary:
\begin{itemize}
    \item We have reasonable confidence that low-impulse glitches are due to rare transients in the interferometer readout and can be kept well under control in ordinary operating conditions.
    \item The fastest force glitches with less-than-a-minute duration may have different explanations, including electronics events, eddy current transients and outgassing events. For all these possibilities, proper ground testing is possible to both consolidate the understanding and reducing the risk that their rate impacts on LISA data quality.
    \item For the long, minute-to-hour force glitches, the only credible explanation that seems to match most of the observational evidence is outgassing. The kind of outgassing events that would explain the observation are somewhat different from the  outgassing spikes observed in the vacuum system under transient conditions (Fig.~\ref{fig:bakeout}). Thus, to consolidate this hypothesis, a dedicated study is needed, and will be performed, also including appropriate experiments.
\end{itemize}

In conclusion, we are confident that the rate of glitches in LISA will be kept to a manageable level, as supported by the indicated tests and studies. The work presented here demonstrates how glitches can be subtracted in the LPF case, which is admittedly simpler than LISA. As with other gravitational wave detectors, identification and mitigation of the remaining glitches will be part of the analysis pipelines that yield LISA’s astrophysical data products. Compared with ground-based gravitational wave instruments, in which both the astrophysical sources and glitches have similar duration, many of LISA’s sources are present in the detector band for much longer timescales than typical glitches, which should aid in distinguishing the two types of signals. Glitch identification and mitigation strategies will still be required to fully realize the LISA science objectives and will be incorporated into global fit algorithms, as described for instance in \cite{PhysRevD.99.024019}. Detailed studies concerning the impact of glitches are ongoing within the LISA Consortium: a dedicated LISA Data Challenge containing LPF glitches \cite{LDC2b} has been released to the community, and initial efforts are already underway to develop and validate glitch mitigation techniques, informed by experience with LISA Pathfinder \cite{PhysRevD.105.042002}.

\section*{Acknowledgments}
This work has been made possible by the LISA Pathfinder mission, which is part of the space-science program of the European Space Agency.\\
We thank Paolo Chiggiato and the vacuum, surfaces and coatings group from Conseil Europ\'{e}en pour la Recherche Nucl\'{e}aire (CERN), for very helpful discussions about the LPF outgassing environment.\\
The Italian contribution has been supported by Istituto Nazionale di Fisica Nucleare (INFN) and Agenzia Spaziale Italiana (ASI), Project No. 2017-29-H.1-2020 ``Attivit\`a per la fase A della missione LISA''. 
The UK groups wish to acknowledge support from the United Kingdom Space Agency (UKSA), the Scottish Universities Physics Alliance (SUPA), the University of Glasgow, the University of Birmingham, and Imperial College London. 
The Swiss contribution acknowledges the support of the Swiss Space Office via the PRODEX Programme of ESA, the support of the ETH Research Grant No. ETH-05 16-2 and the support of the Swiss National Science Foundation (Projects No. 162449 and No. 185051).  
The Albert Einstein Institute acknowledges the support of the German Space Agency, DLR. The work is supported by the Federal Ministry for Economic Affairs and Energy based on a resolution of the German Bundestag (No. FKZ 50OQ0501, No. FKZ 50OQ1601, and No. FKZ 50OQ1801). 
J.I.T. and J.S. acknowledge the support of the U.S. National Aeronautics and Space Administration (NASA). 
The Spanish contribution has been supported by Contracts No. AYA2010-15709 (Ministerio de Ciencia e Innovaci\'on MICINN), No. ESP2013- 47637-P, No. ESP2015-67234-P, No. ESP2017-90084-P (Ministerio de Asuntos Econ\'omicos y Transformaci\'on Digital, MINECO) and No. PID2019-106515GB-I00 (MICINN). Support from AGAUR (Generalitat de Catalunya) Contract No. 2017-SGR-1469 is also acknowledged. M.N. acknowledges support from Fundacion General CSIC (Programa ComFuturo). F.R. acknowledges an FPI contract from MINECO. 
The French contribution has been supported by the CNES (Accord Specific de projet No. CNES 1316634/CNRS 103747), the CNRS, the Observatoire de Paris and the University Paris-Diderot. E.P. and H.I. would also like to acknowledge the financial support of the UnivEarthS Labex program at Sorbonne Paris Cit\'e (No. ANR-10-LABX-0023 and No. ANR-11-IDEX-0005-02).
N.K. would like to thank for the support from the CNES Fellowship.

 \appendix
 \section{\label{app:AEE} ESTIMATE OF PARAMETER ERRORS}
 
 As the fitting procedure in Sec.~\ref{sec:gltchdetection}  is non optimal and non linear, we estimated the fitting parameter covariance from the Cram\'er-Rao bound \cite{kay2013fundamentals}, assuming the noise is Gaussian, and  using the measured PSD of the residuals. For glitches in ordinary runs, we have checked that an optimal filter procedure, for which the Cram\'er-Rao bound becomes an exact estimate, returns parameter values that are in agreement, within the estimated uncertainty, with those found with our non optimal method. 

To this aim, once a glitch had been identified and fitted, we have: expanded  the fitting function around the best fit parameter values, up to linear terms in the fitting parameters; applied  the optimal linear filter method for multi-component signals \cite{KL};  calculated the fitting amplitudes; finally propagated these results  back to that of the original fitting parameters. The results of both  procedures agree within the uncertainty estimated from the Cram\'er-Rao bound, except for a few outliers. 

We were not able to apply the optimal procedure also to the cold runs data, because of their above mentioned complexity. Thus, for consistency, in the following  we use, for both ordinary and cold runs, the parameter values resulting from the nonoptimal, time domain procedure.

Our error estimates represent certainly a lower bound. In particular, given the length of the low pass filter we use for glitch identification, the uncertainty on $\Delta$ for the shortest glitches, $\Delta \lesssim \SI{30}{s}$, may be significantly underestimated. However we stress that  none of the results depends critically on the accuracy of such an uncertainty estimate, as parameter fluctuations within the glitch population are significantly larger than their uncertainties.
 
 \section{\label{app:RateLikelihood} ESTIMATION OF GLITCH RATE}
The rate of impulse-carrying glitches in both ordinary runs and subsets of cold runs is computed with a Bayesian analysis, as follows. According to the Lilliefors test, the distribution of the waiting times $t_i$ is compatible with an exponential distribution. We can then provide an estimate of the rate $\lambda$. Under the assumption that each event is an independent random extraction from an exponential distribution with rate $\lambda$, we apply Bayes' theorem to the joint probability of $n$ events, with a uniform prior on $\lambda$, to get its posterior distribution:
\begin{equation}
    \mathcal{L}(\lambda) = \frac{S^{n+1}}{\Gamma{(n+1)}} \lambda^n e^{-S \lambda}, \quad S=\sum_i t_i
\end{equation}
where $\Gamma$ is the Gamma function. We estimate $\lambda$ as the probability-maximizing value, and the asymmetric error bounds at 1$\sigma$ confidence level as the quantiles of the probability distribution containing 68\% of it. The result is compatible with the fit of the waiting time distributions, in Fig.~\ref{fig:tauhistograms}.
 
 \section{\label{app:APCR} PROJECTION OF GLITCH COUNT RATE}
 
 The  projection of the glitch count rate from ordinary to cold runs has been calculated as follows. From the observed counts in ordinary runs, we made a Bayesian estimate of the posterior distribution for the probabilities of the three different glitch categories. In addition we assumed  a Poisson distribution for the  number of glitches during cold runs, with the posterior distribution for the rate derived from the  ordinary runs as explained in  Sec.~\ref{sec:glitchparamstats}. We then integrated the probabilities  for the counts in the three categories, as calculated from the proper multinomial distribution,  over all the posteriors above, to obtain the total probability for those same counts.

\bibliography{bibliography.bib}


\end{document}